\documentclass[10pt, journal, compsoc]{IEEEtran}

\usepackage{mystyle}
\usepackage{listings}

\newcommand{\commentblock}[1]{}

\togglefalse{complete-version}
\settoggle{highlight-revision}{true}

\def\doctitle{Multi Language Models for On-the-Fly Syntax Highlighting}

\def\docauthors{Marco Edoardo Palma, Harald C. Gall}

\def\dockeywords{%
Syntax highlighting, neural networks, deep learning, regular expressions
}

\StrSubstitute{\doctitle}{\\}{ }[\cleandoctitle]
\StrSubstitute{\dockeywords}{.}{}[\cleandockeywords]
\hypersetup{
	pdftitle={\cleandoctitle},
	pdfauthor={\docauthors},
	pdfkeywords={\cleandockeywords}
}

\DeclareAcronym{api}{
	short = API,
	long = {Application Program Interface}
}

\DeclareAcronym{bert}{
	short = BERT,
	long = {Bidirectional Encoder Representations from Transformers}
}

\DeclareAcronym{ast}{
	short = AST,
	long = {Abstract Syntax Tree}
}

\DeclareAcronym{cfg}{
	short = CFG,
	long = {Control Flow Graph}
}

\DeclareAcronym{gpt}{
	short = GPT,
	long = {Generative Pretrained Transformer}
}

\DeclareAcronym{ir}{
	short = IR,
	long = {Information Retrieval}
}

\DeclareAcronym{lstm}{
	short = LSTM,
	long = {Long Short-Term Memory}
}

\DeclareAcronym{nn}{
	short = NN,
	long = {Neural Network}
}

\DeclareAcronym{rnn}{
	short = RNN,
	long = {Recurrent Neural Network}
}

\DeclareAcronym{brnn}{
	short = BRNN,
	long = {Bidirectional Recurrent Neural Network}
}

\DeclareAcronym{cnn}{
	short = CNN,
	long = {Convolutional Neural Network}
}

\DeclareAcronym{sota}{
	short = SOTA,
	long = {state-of-the-art}
}

\DeclareAcronym{tn}{
	short = TN,
	long = {Token Normalization}
}

\DeclareAcronym{tf-idf}{
	short = tf-idf,
	long = {term frequency–-inverse document frequency}
}

\DeclareAcronym{anova}{
	short = ANOVA,
	long = {ANalysis Of VAriance}
}

\DeclareAcronym{ta}{
	short = TA,
	long = {Task-Adaptive}
}

\DeclareAcronym{gru}{
	short = GRU,
	long = {Gated Recurrent Unit}
}

\DeclareAcronym{sh}{
	short = SH,
	long = {syntax highlighting}
}

\DeclareAcronym{re}{
	short = regex,
	short-plural = es,
	long = {regular expression}
}

\DeclareAcronym{om}{
	short = OM,
	long = {Oracle Methods}
}

\DeclareAcronym{eta}{
	short = ETA,
	long = {Extended Token Annotation}
}

\DeclareAcronym{heta}{
	short = HETA,
	long = {Highlighted Extended Token Annotation}
}

\DeclareAcronym{ide}{
	short = IDE,
	long = {Integrated Development Environment}
}

\DeclareAcronym{bf}{
	short = BF,
	long = {brute-force}
}

\DeclareAcronym{peg}{
	short = PEG,
	long = {Parsing Expression Grammar}
}

\DeclareAcronym{dl}{
	short = DL,
	long = {deep learning}
}

\DeclareAcronym{singlelang}{
	short = SL,
	long = {Single Language}
}

\DeclareAcronym{multilang}{
	short = ML,
	long = {Mutli Language}
}

\DeclareAcronym{da}{
	short = DA,
	long = {Deep Abstraction}
}

\begin{document}

\title{\doctitle}

\author{
    Marco Edoardo Palma, Pooja Rani, Harald C. Gall

    \IEEEcompsocitemizethanks{ %
        \IEEEcompsocthanksitem The authors are with the University of
        Zurich, Zurich, Switzerland. E-mail:
        \href{mailto:marcoepalma@ifi.uzh.ch}{marcoepalma@ifi.uzh.ch},
        \href{mailto:rani@ifi.uzh.ch}{rani@ifi.uzh.ch},
        \href{mailto:gall@ifi.uzh.ch}{gall@ifi.uzh.ch}.
    }
}

\IEEEtitleabstractindextext{
    \begin{abstract}
Syntax highlighting is a critical feature in modern software development
environments, enhancing code readability and developer productivity. However,
delivering accurate highlighting in real-time remains intractable in online and
web-based development tools due to strict time and memory constraints on backend
services. These syntax highlighting systems must serve highlights rapidly and frequently, including
in scenarios where code is only partially valid or entirely invalid. This has
led to the concept of on-the-fly syntax highlighting; where visual annotations
are generated just before content is served online, often at high request rates
and under incomplete input conditions.
To meet these demands efficiently, state-of-the-art models leverage
Convolutional Neural Networks to automatically learn the behavior of
brute-force syntax highlighting resolvers; tools that are easy for developers to
implement but too slow for deployment. Through a process we refer to as Deep
Abstraction, these brute-force strategies are encoded into fast, statistical
models that offer both high accuracy and low-latency inference. Despite their
success, such models still face key challenges: they are limited to supporting a
single programming language per model, require the generation of large datasets via slow
and inefficient brute-force generators, and involve long and resource-intensive
training sessions. In multi-language environments, this leads to the need for
maintaining and deploying multiple independent models, one per language, which
increases system complexity and operational overhead.
This work addresses these challenges by introducing a unified model capable of
effectively highlighting up to six mainstream programming languages, thereby
reducing deployment complexity by a factor of six and improving performance on
previously unseen languages. A novel normalization technique is proposed, which
significantly enhances model generalization to languages it has not been
explicitly trained on. Furthermore, the study explores few-shot learning tasks
aimed at reducing the cost of training syntax highlighting models. By relying on
a small number of manually generated oracle samples instead of large datasets,
this approach minimizes dependence on brute-force highlighters and reduces
training effort. The proposed normalization step further boosts model accuracy
under these constraints, paving the way for efficient, scalable, and
cost-effective syntax highlighting across a wide range of programming languages.
\end{abstract}


}
\maketitle

\section{Introduction}
\label{sec:introduction}

\Ac{sh} involves visually annotating code by applying distinct colours to specific language sub-productions,
thereby enhancing code readability and comprehension~\cite{sarkar_impact_2015, 10.1145/2858036.2858372}.
This feature is standard in most modern \ac{ide} and is widely employed in various online contexts, such as
code review platforms, repository file browsers, and code snippet displays, all benefiting from \ac{sh}
mechanisms~\cite{myPaper1}. Notably, various online platforms perform \emph{dynamic}, or \emph{On-the-Fly}, \ac{sh},
meaning that \ac{sh} resolvers compute the highlighting for code immediately before displaying it to the user.

The choice of dynamic \ac{sh}, influenced by limited storage of file or snippet's metadata and the challenges
of caching notations, presents significant technical demands on \ac{sh} resolvers. While running client software on users’ machines (such as in browsers)
could theoretically handle this task, it is generally avoided due to the extensive computational resources required.
These resolvers must operate efficiently under high request volumes, ensuring platform usability by maintaining scalability and fast response times.
Additionally, they must deliver accurate highlighting by reliably associating code sub-productions with the appropriate \ac{sh} classes or colours.

However, achieving this level of accuracy (defined as the percentage of
non-whitespace tokens correctly classified into their corresponding \emph{SH} class)
within tight time and resource constraints is
challenging, as it requires the resolvers to perform a grammatical analysis of
the code being highlighted. A full parsing process is often impractical due to
time limitations and the risk of being unable to parse incorrect language derivations~\cite{myPaper1}.
The need for rapid development also persists, given the rapid evolution of mainstream
programming languages and their versions. These factors help explain why developers experience substantially poorer
syntax highlighting online than in local environments~\cite{myPaper1}. Historically, developers have manually constructed intricate
systems of regular expressions to accomplish \ac{sh}; a method effective but prone to tedium and
inaccuracies~\cite{pygments, myPaper1}. Specifically, developers must derive complex regular expression systems for
each language to identify and colour code sub-productions, representing lexical or grammatical roles
(e.g., integer literal, function identifier).
This highlights a strong need for syntax highlighters that are fast, scalable, and responsive, whilst also
being accurate, adaptable to evolving language features, and reboust across diverse programming languages.

The current \ac{sota} approach addresses these goals by treating \ac{sh} as a machine learning translation
problem~\cite{myPaper1, myPaper2} through a process known as \ac{da}.
\Ac{da} involves generating a fast statistical resolver for tasks that have easily derived \ac{bf} algorithms but
lack efficient solutions, automating the creation of an efficient statistical resolver for the \ac{bf} model.
Practically, developers first produce a basic brute force resolver for \ac{sh}, which a statistical model then
optimizes, adding robustness against invalid language derivations. This approach enables developers to design a
deterministic \ac{ast} walker for each language, creating a \ac{bf} \ac{sh} resolver that excels in accuracy and
grammatical coverage but is unsuitable for \emph{On-the-Fly} scenarios due to large prediction delays and
inconsistent performance. Consequently, the \ac{bf} model is used to generate an oracle dataset consisting of
mappings between valid language derivations and \ac{sh} tags (colour abstractions) for each token. A statistical
model is subsequently trained to generate \ac{sh} tags for any given language derivation, resulting in resolvers
that maintain the \ac{bf} model’s high accuracy on both valid and invalid derivations while significantly
reducing prediction delays.

However, this current approach has two key limitations. First, generating the oracle dataset requires at least 13,000
samples—a substantial demand, considering the inefficient \ac{bf} models must produce \ac{sh} output for each
sample, with \ac{cnn}-based \ac{sota} models requiring four training passes over this dataset.
Additionally, the resulting statistical \ac{sh} models are \ac{singlelang}, meaning they support only one programming
language. Integrators must therefore create a separate \ac{bf} model, produce a 13,000-sample oracle, and train
and deploy individual statistical highlighters for each language.
The \ac{singlelang} nature of \ac{sota} models is particularly limiting, as these models cannot
highlight languages they were not trained on. In contrast, \emph{state-of-practice} resolvers
support hundreds of languages~\cite{pygments, treesitter}. This makes it essential to carry out
these training processes.

To reduce the costs and complexities associated with creating and deploying such \emph{On-the-Fly}
\ac{sh} models in multi-language environments, this work introduces \ac{multilang} models for
\ac{sh}. \Ac{multilang} models can encode the \ac{sh} behaviour of at least six \ac{bf} models.
Furthermore, by implementing a novel input normalization strategy, this work demonstrates how these
models can improve \ac{sh} performance on mainstream programming languages. Lastly, this research
explores how the normalization strategy can substantially reduce the incremental number of oracle
samples required to adapt an pre-trained model to additional languages, lowering this adaptation
cost from 13,000 samples per language to as few as 10. The results evaluate \ac{sh} accuracy on both
valid and invalid language derivations, comparing each model’s accuracy against the best-performing
single-language models, thereby assessing the potential of multilingual models to complement or, in
some cases, replace specialized single-language models in this domain.

The implementation, new multi-language and few-shot benchmark datasets, and
results are available in the replication package~\cite{replicationpackage}.

The rest of the paper is structured as follows: Section~\ref{sec:approach} outlines the
requirements for multilingual models in this domain and introduces the Token
Normalization strategy; Section~\ref{sec:experiments} presents the research
questions, the construction of the multilingual datasets, and the training and
validation tasks; Section~\ref{sec:results} analyzes the findings for each
research question; Section~\ref{sec:related_work} surveys the related literature;
and Section~\ref{sec:conclusions} summarizes the
contributions and insights, and discusses future directions in this area.

\section{Approach}
\label{sec:approach}
In the development of multi-language models for \emph{on-the-fly} \ac{sh}, two key
challenges arise. First, the original single-language models cannot easily
generalize their learned highlighting patterns to new languages. Second, each
language uses a distinct set of token IDs, causing identical syntactic elements
to appear as disjoint integer streams. The proposed approach addresses both of
these issues: it outlines how to adapt existing \ac{sota} deel learning models
to multi-language \ac{sh} and then proposes a \ac{tn} that consolidates token
types across languages to boost model accuracy in multi-language scenarios.

\subsection{Multi-Language Syntax Highlighting Models}
This work builds upon the \ac{sota} strategies and models for
\emph{on-the-fly} \ac{sh}. Currently, the \ac{sota} resolvers for this task
are \ac{cnn}-based models that approximate the behavior of \ac{bf}
\ac{sh} algorithms. These BF algorithms, developed specifically for
each language, leverage formal grammar parsing to derive the \ac{ast} and assign
each language-specific token to one of 12 \ac{sh} classes~\cite{myPaper1}. These classes fall
within four broader grammatical macrogroups: \emph{Lexical}, \emph{Identifier},
\emph{Declarator}, and \emph{Annotation}.

Despite their accuracy, \ac{bf} approaches are impractical for
\emph{on-the-fly} scenarios due to high computational costs and the inability to
process incomplete or syntactically incorrect derivations. To address this
limitation, prior work~\cite{myPaper1, myPaper2} introduced the \ac{da} process, which
automatically compiles \ac{bf} algorithms into statistical models. These
models achieve near-perfect \ac{sh} accuracy while significantly reducing evaluation
time and maintaining the same levels of accuracy also on invalid language derivations.

The current \ac{sota} models employ \ac{cnn} architectures
tailored for \ac{sh} in a single language. Each model consists of an
embedding layer ($\text{Emb}: \mathbb{N}^{vocab_size} \rightarrow
\mathbb{R}^{32}$), followed by two convolutional layers ($C_i: \mathbb{R}^{32}
\rightarrow \mathbb{R}^{32}$) activated by ReLU ($\sigma$), processing input
sequences bidirectionally. Dropout regularization ($\delta(p=0.3)$) is applied
to these layers to mitigate overfitting. The concatenated outputs ($\oplus$) are
passed to a third convolutional layer ($C_3: \mathbb{R}^{2*32} \rightarrow
\mathbb{R}^{256}$), and the extracted features are classified via a fully
connected feedforward layer ($FC: \mathbb{R}^{256} \rightarrow \mathbb{N}^{12}$)
into the respective highlighting classes \emph{hc}. These models have
established the benchmark for \ac{sh}~\cite{myPaper2}, achieving the highest
accuracy across valid and invalid derivations with minimal inference time. Three
variations, \emph{CNN32}, \emph{CNN64}, and \emph{CNN128}, differing in
embedding and hidden dimensions, were identified as best-performing
configurations with negligible accuracy variance.

This work extends these \ac{sota} models to support multi-language syntax
highlighting without altering the architecture in ways that would degrade
the prediction delays. The primary motivation is to maintain efficient inference while
enabling a single model to process multiple programming languages, thereby
reducing deployment overhead.

A key challenge in designing multi-language models is the increased vocabulary
size as a result of considering all the language features of more than one language.
The \ac{da} approach relies on \ac{cnn} models receiving
token IDs assigned by the language’s lexer—unique integer values representing
lexical components. Since each programming language defines a different set of
token types, a single model must accommodate variations in token vocabulary
across multiple languages; such as extending a \ac{sh} for \java to support \python,
the highlighter must be expanded to recognize \python -specific language tokens such as
strong keywords like \emph{def}, indentation tokens, or string interpolation parts.
To address this, the architecture is adapted to follow the design proposed by
Palma et al., with one modification: a standardized input dimension large enough
to support the token types from all targeted languages. Unlike previous models,
where input size was tailored to a specific language, the multi-language model
requires a unified input structure capable of handling multiple lexers’ token
outputs. This adjustment ensures scalability while preserving the efficiency and
accuracy of the underlying CNN-based \ac{sh} approach.

\subsection{Token Normalization}
\ac{da} models for \ac{sh} operate on language-specific lexical
token IDs.
These token ID sets vary significantly across languages, preventing direct generalization of learned
highlighting patterns from one language to another. Even when languages share
common grammatical structures, a model trained on one language is unable to
recognize the same patterns in another due to differences in token ID
assignments. This constraint is a primary reason why current \ac{sota} \ac{sh}
models are single-language only, requiring a separate model for each new
programming language.

For instance, consider the task of identifying class declarator identifiers,
such as the token \emph{Payment} in the Java derivation: \code{class Payment \{\}}.
In this case, \emph{Payment} is highlighted as a \emph{class\_declarator} within
the \emph{Declarator} macrogroup. The model receives token sequences generated
by the Java lexer, which might take the form \{10, 102, 45, 46\}, where 10
represents \code{class}, 102 corresponds to an identifier (\code{Payment}), and
45 and 46 denote the opening and closing braces. The model, during training,
learns that a token 102 preceded by a token 10 should be classified as a
\emph{class\_declarator}. However, if the same model were applied to the
equivalent C\# derivation: \code{class Payment \{\}}
the \emph{C\#} lexer would produce a completely different token sequence, preventing the
model from recognizing the pattern. This discrepancy, which occurs across
programming languages, limits the generalizability of \ac{sh} models and hinders
their deployment in multi-language settings.

Two potential solutions were considered to address this limitation. The first
approach involves designing a universal lexer to generate consistent token
sequences across languages. While this would ensure uniformity in dataset
generation and model training, it is impractical due to the need for extensive
parser modifications. The second approach proposes using a universal lexer
exclusively for multi-language models, tokenizing program text into a new,
language-independent set of tokens. However, this method suffers from potential
incompatibilities in tokenization rules, particularly for features such as
string interpolation, leading to accuracy degradation.
To overcome these challenges, this work introduces the \ac{tn}, a component that
normalizes token types across languages. The \ac{tn} maps equivalent lexical
elements, such as \code{+} or \code{[a-zA-Z]+}, to a fixed token type, ensuring consistency
despite differences in token IDs assigned by language-specific lexers. Tokens
not covered by a normalization rule are passed to the model in their original
form. As a result, the model receives identical token sequences for
syntactically equivalent constructs across different languages, allowing it to
generalize its learned highlighting patterns.
The \ac{tn} operates as a lookup mechanism applied before the model’s input
processing, incurring no additional computational cost in terms of time or space
complexity. This ensures that model performance remains unaffected while
significantly enhancing its ability to handle multi-language \ac{sh} tasks.
Pseudo-code in Listings~\ref{lst:tn} shares the routines required for both building the normalization
bindings from a set of languages, and later infer the normalization or offsets during
inference. Further implementation details used in this work are available in the
replication package~\cite{replicationpackage}.
\emph{TN} is designed to map syntactically equivalent token categories across multiple programming
languages into a shared identifier space, while preserving language-specific tokens through disjoint
offsets. The normalization process operates in two stages: construction of shared token bindings and
inference-time normalization.
During construction, the token definitions of all supported languages are analyzed. Each token is
described by a rule or regular expression defining the syntactic pattern it matches. Tokens whose
definitions are shared by at least two languages, either through exact or semantically equivalent
rules, are identified as shared tokens. These shared tokens are assigned normalized identifiers in a
contiguous shared region starting at index 0. This shared region enables the model to learn
cross-language syntactic patterns, such as identifiers, literals, keywords, or comment structures,
using a single representation.
Tokens that do not appear in this shared region are treated as language-specific. To avoid
collisions between these tokens and the shared identifiers, each language is assigned a unique
offset immediately following the shared region. Language-specific token identifiers are then shifted
by their corresponding offset at inference time. As a result, all tokens, whether shared or
language-specific, are mapped into a single fixed identifier space while maintaining strict
separation between shared and non-shared regions.
The dimensionality of the input space is therefore determined by the size of the shared region plus
the maximum number of token identifiers required to accommodate the largest language-specific token
set. In this work, a dimensionality of 315 corresponds to the maximum identifier index needed to
represent all shared tokens and all offset-shifted language-specific tokens without overlap. This
value is not arbitrary, but rather the minimal dimension that guarantees collision-free
normalization across all languages considered. To ensure architectural consistency and avoid bias,
all models, including single-language baselines, are projected into this same 315-dimensional space,
even when TN is not applied.
At inference time, tokens that match shared normalization rules are mapped directly to their
normalized identifiers in the shared region. Tokens that do not match any shared rule are passed
through by applying the language-specific offset, preserving their identity while remaining
compatible with the fixed input representation. This mechanism ensures deterministic behavior, full
coverage of all token types, and reproducibility across languages and experimental configurations.

While this approach facilitates the transfer of shared highlighting patterns across languages, it is
not intended to enforce identical highlighting decisions for all syntactic constructs. Programming
languages inherently differ in their grammar and syntactic sugar, and certain language-specific
constructs must therefore be handled distinctly by design. \emph{TN} explicitly preserves
these differences by leaving language-specific tokens unnormalized and allowing the model to learn
their behavior independently. As such, differences in highlighting for these constructs are not
errors, but a necessary consequence of accurate, language-aware \emph{SH}.

At the same time, \emph{TN} enhances the model’s adaptability by substantially reducing the number of
language-specific patterns that must be learned during multi-language training. Shared grammatical
sub-constructs, such as identifiers, literals, and common structural elements, are normalized across
languages, enabling effective transfer of highlighting behavior. The remaining non-shared patterns
are inferred from the presence of language-specific tokens in the input sequence, allowing the model
to correctly condition its predictions on the target language.

\begin{figure*}[!t]
\centering
\includegraphics[width=\textwidth]{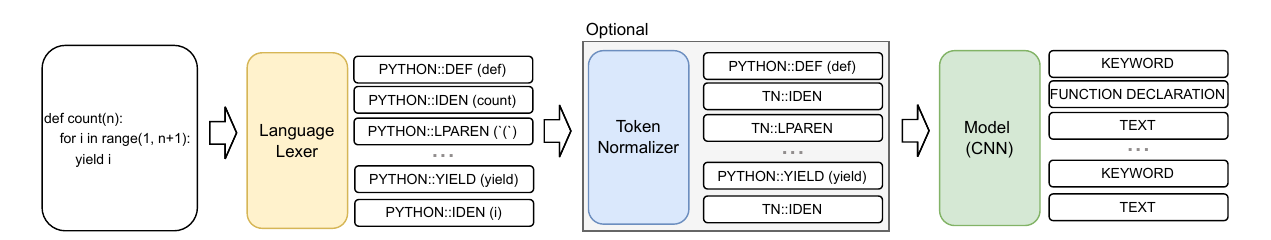}
\caption{
Overview of the syntax highlighting synthesis pipeline. A plain-text language derivation is first
processed by a language-specific lexer, which converts the input into a sequence of
language-specific tokens. Optionally, these tokens are passed through a constant-time token
normalizer that maps shared token categories (e.g., identifiers and literals) into a shared integer
space while preserving language-specific tokens via separate mappings. The resulting token sequence
is then processed by a statistical model that assigns a synthesizing type to each token, which can
subsequently be translated into concrete visual attributes by the application.
}
\label{fig:flowdiagram}
\end{figure*}

This design is particularly beneficial in two scenarios: (1) when introducing support for a new or
previously unseen programming language, and (2) when training multi-language models with a limited
number of samples per language. In these cases, the model does not require exhaustive exposure to
the full grammar of each language, but instead needs to learn only the language-specific exceptions
and unique constructs, significantly reducing training overhead while preserving high highlighting
accuracy.
A high-level view of the data flow and the impact of introducing the \emph{TN} into the
pipeline is illustrated in Figure~\ref{fig:flowdiagram}, which highlights how a language derivation
is eventually annotated.

\begin{lstlisting}[
  caption={
  Pseudo-code for the Token Normaliser: algorithm to construct per-language normalization bindings for shared tokens, and the normalization procedure applied at inference time.
  },
  label={lst:tn},
  basicstyle=\ttfamily\footnotesize,
  frame=single
]
// Inputs:
//  defs[l] = token definitions (id, regex/rule)
//            for language l
// Outputs:
//  bind[(l,id)] -> normalized id for shared tokens
//  off[l]       -> offset for non-shared tokens
// Shared tokens use ids [0..n-1] or are shifted.
FUNCTION BUILD_TN(defs_by_lang):
  // Count how many languages define each regex.
  count = Map()   // regex -> int
  FOR EACH l IN defs_by_lang:
    FOR EACH d IN defs_by_lang[l]:
      // Group by exact or semantic match.
      count[d.regex] = count.get(d.regex, 0)+1
  // Assign normalized token ids.
  shared = Map()  // regex -> norm_id
  n = 0
  FOR EACH rx IN count:
    IF count[rx] >= 2:
      shared[rx] = n
      n = n + 1
  // Build per-language bindings for shared tokens
  bind = Map()    // (l, id) -> norm_id
  FOR EACH l IN defs_by_lang:
    FOR EACH d IN defs_by_lang[l]:
      IF shared.CONTAINS(d.regex):
        bind[(l, d.id)] = shared[d.regex]
  // Offsets for non-shared tokens.
  off = Map()     // l -> offset
  base = n
  FOR EACH l IN defs_by_lang:
    off[l] = base
    base = base + (1 + MAX_ID(defs_by_lang[l]))
  RETURN (n, bind, off)
FUNCTION NORMALIZE(l, token_ids, bind, off):
  out = []
  FOR EACH id IN token_ids:
    IF bind.CONTAINS((l, id)):
      out.APPEND(bind[(l, id)])  // Normalized
    ELSE:
      out.APPEND(off[l] + id)    // Shifted
  RETURN out
\end{lstlisting}

\section{Experiments}
\label{sec:experiments}

This study evaluates the performance of \ac{sh} models in
multi-language tasks. Although \ac{sota} models perform
exceptionally well in single-language domains~\cite{myPaper2}, their reliance on extensive
retraining to accommodate new languages presents scalability challenges. To
address this limitation, the research explores the potential of multi-language \ac{sh}
models, comparing their effectiveness against \ac{sota} single-language models. It
also investigates the role of \ac{tn} in enhancing performance
and assesses model capabilities in few-shot scenarios, where oracle samples are
limited to a small number. By analyzing these aspects, the study aims to
identify strategies that improve the efficiency and accuracy of \ac{sh} tasks. The
research questions framing this investigation are outlined below.

\begin{figure}[!t]
\centering
\includegraphics[width=\columnwidth]{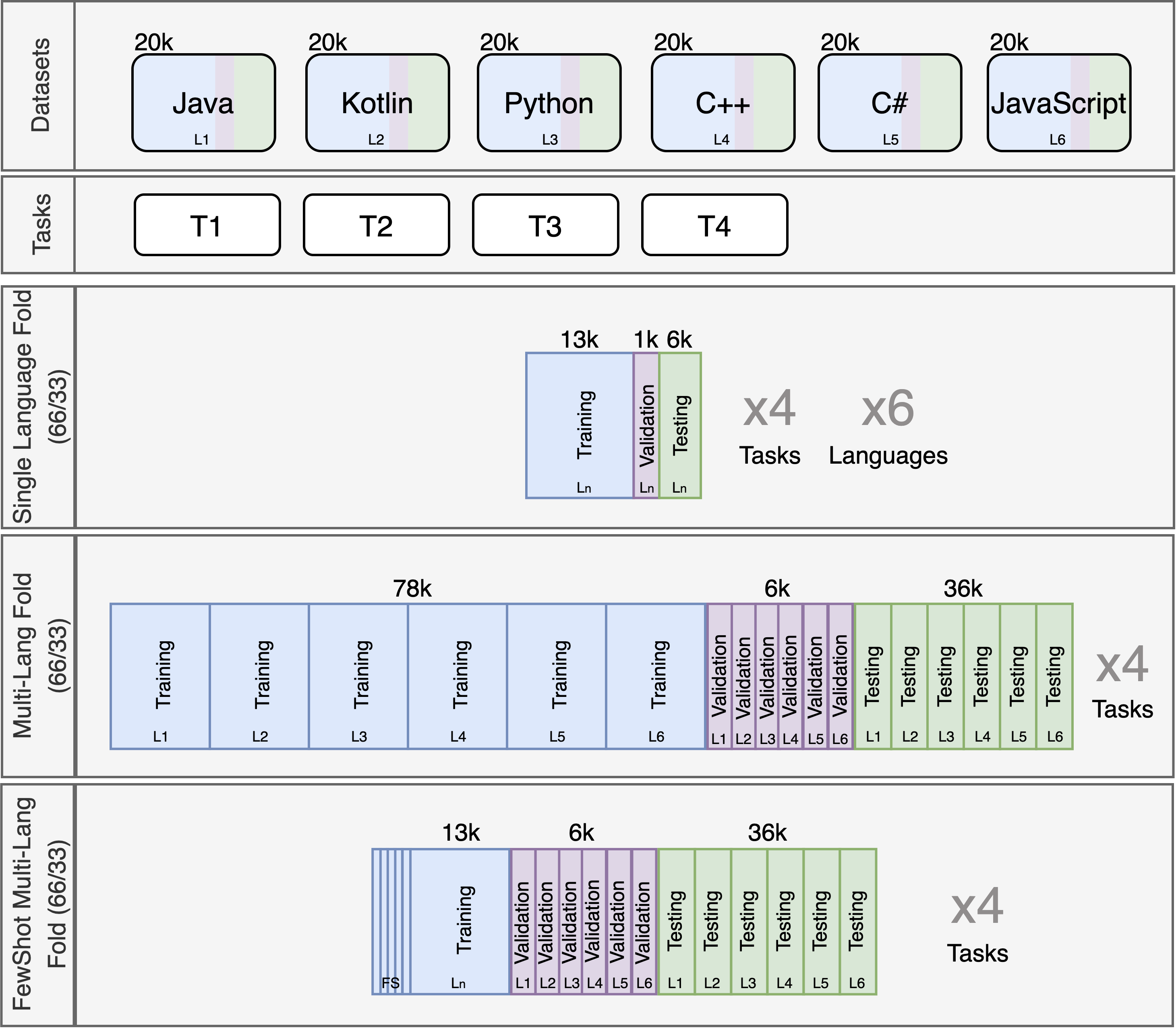}
\caption{
Illustration of how the original dataset of 20k samples per language is
structured for training and validation within a single fold:
\emph{Single-Language Task} A CNN model is trained on a single language and
tested on its respective test set, repeated for each coverage task and language.
\emph{Multi-Language Task}: A model is trained on a merged dataset of all six
languages and evaluated on each language’s test set, repeated for each coverage
task. \emph{Few-Shot Task}: A model is trained on a single language (Ln),
fine-tuned on a small sample from other languages (FS), and tested on the same
test sets as the other tasks.
}
\label{fig:trainscenariosfold}
\end{figure}

\begin{reqs}
\item [\req{1}] How do state-of-the-art syntax highlighting models perform on
unseen mainstream programming languages, and how does Token Normalization
influence their ability to natively generalize to new languages?
\end{reqs}
Current \ac{sota} models are typically tailored to individual languages and
require significant retraining to accommodate new languages. Token
Normalization, by mapping shared token types across languages to unified
representations, aims to enhance the generalization capability of SH models.
This study evaluates the out-of-the-box performance of \ac{sota} models on unseen
languages and quantifies the accuracy improvements
achieved through \ac{tn}.
\smallskip

\begin{reqs}
\item [\req{2}] How do multi-language syntax highlighting models compare to
state-of-the-art single-language models when applied to mainstream programming
languages?
\end{reqs}
While \ac{sota} \ac{sh} models have primarily focused on individual languages,
this study investigates whether CNN-based architectures can be effectively
trained on datasets encompassing multiple languages. The goal is to compare
their performance with single-language models and assess the feasibility of
multi-language training without sacrificing accuracy.
\smallskip

\begin{reqs}
\item [\req{3}] How do multi-language syntax highlighting models, fine-tuned on
few-shot datasets, perform compared to multi-language models and the
state-of-the-art single-language models?
\end{reqs}
This question explores the adaptability of multi-language models trained on a
limited number of samples from new languages, compared to the currently required large
training datasets of 13k samples per language. It seeks to determine whether
fine-tuning on a few samples allows these models to approach or exceed the
performance of models trained on extensive datasets, potentially making
multi-language training more efficient.
\smallskip

\begin{reqs}
\item [\req{4}] How does Token Normalization affect the performance of
multi-language syntax highlighting models trained on few-shot datasets?
\end{reqs}
During few-shot training allows deep learning models to benefit from only a
reduced number of samples compared to \ac{sota} processes, which require a
large corpus of training samples (13k). This means that during few-shot training
the models may have less evidence of \ac{sh} patterns for each language, which
may lead to a reduction in \ac{sh} accuracy.
By design, \ac{tn} can be applied to any model in this context, and aims to
reduce the problem space of multi-language tasks by mapping the same language
token types to the same numerical representation. This means that if the same
highlighting pattern is found in more than one language, the model only needs to learn
this pattern once to generalize it to the other languages.
Given the multi-language problem and the goals of the \ac{tn} introduced here,
this research question examines whether it can further improve the
accuracy of multi-language models when they are trained with limited samples.
\smallskip

\subsection{Datasets}
\commentblock{
This work builds on top of the \ac{sh} dataset developer by Palma \etal \cite{myPaper2},
and extends it to add multi-language and few shot learning task. the following section describes
the content of this baseline dataset, how a multi-language dataset is generated, and how few shot
learning datasets are generated.
}
This work builds upon the \ac{sh} dataset developed by Palma \etal
\cite{myPaper2}, extending it to support multi-language tasks and few-shot
learning scenarios. The following section outlines the contents of the baseline
dataset, details the process of generating multi-language datasets, and explains
the methodology for creating few-shot learning datasets.

\paragraph{Baseline dataset}
This study utilizes the \ac{sh} dataset which provides \ac{sh} for six mainstream programming
languages: \java, \kotlin, \python, \cpp, \csharp, and \javascript~\cite{myPaper2}. The dataset comprises
20,000 unique language derivations per language, with no syntactic duplicates,
generated using manually developed brute-force syntax highlighting resolvers. It
has been previously employed to train and evaluate state-of-the-art models for
on-the-fly syntax highlighting.

The baseline dataset contains mappings between sequences of language derivation tokens
and their corresponding 12 \ac{sh} classes. Tokens are represented by their token
IDs—integer values assigned by the original lexer of the respective language.
Similarly, syntax highlighting classes are encoded as integer values
corresponding to categories: \emph{keyword}, \emph{literal}, \emph{char\_string\_literal},
\emph{comment}, \emph{type\_identifier}, \emph{function\_identifier}, \emph{field\_identifier},
\emph{class\_declarator}, \emph{function\_declarator}, \emph{variable\_declarator}, and
\emph{annotation\_declarator}. These classes align with \emph{Coverage Task 4}, which is the most
comprehensive syntax highlighting task, encompassing all lexical and grammatical
token types: Lexical, Identifier, Declarator, and Annotation groups.
The dataset allows for the conversion to the other three \emph{Coverage Task} (1, 2, and 3)
through the \emph{Task Adapter} \cite{myPaper1}.

Additionally, the dataset includes oracles for incomplete or invalid language
derivations, enabling the validation of \ac{sh} models on code snippets. The
invalid derivations dataset follows the same format as the valid derivations and
is crafted to reflect language-specific snippet lengths based on mean, standard
deviation, minimum, and maximum line numbers obtained from \emph{StackExchange}
data~\cite{stackexchange}.

This dataset is the most comprehensive currently available for evaluating syntax
highlighting tools against a fully accurate and deterministic ground truth. It
has been extensively used for training and validating \ac{sota} models
and comparing their performance with popular SH tools such as \pygments~\cite{pygments} and
Tree-sitter~\cite{treesitter}. Furthermore, it includes predefined three-fold splits for valid and
invalid derivations, with a 33\%-66\% division into testing and training sets, and
10\% of the training data reserved for validation. For each fold, 5000 incorrect
derivations are generated from the test subset, ensuring robust validation of \ac{sh}
models across various scenarios.

\paragraph{Multi Language Dataset}
The multi-language dataset is constructed by restructuring the baseline dataset
\cite{myPaper2}. For each of the three folds across the six languages, the
training datasets are combined and shuffled into a single multi-language training
dataset. This process creates three cross-validation folds, each with a
consolidated multi-language training dataset, while preserving the original
per-language validation, test, and snippet datasets for each fold.  This
approach ensures a multi-language dataset with no duplication between the
training and test sets, enabling a direct per-language comparison of
multi-language models with the \ac{sota} single-language models.

\paragraph{Multi Language Few Shot Dataset}
\commentblock{
This work extends this dataset by including few shot learning tasks.
These tasks are created by creating subsets of each fold of each language's training dataset by randomly sampling
with replacement a number of samples until the target few shot learning samples is reached.
The result is the addition of 5 new alternative training dataset to each of the fold of each language.
test, validation, and snippet test datasets are preserved allowing for a direct comparison of multi-language
models trained on few shot learning tasks to each \ac{sota} single language model.
}
This study extends the dataset to include few-shot learning tasks. These tasks
are generated by creating subsets of each fold of each language's training
dataset. This is done by randomly sampling, with replacement, until the desired
number of few-shot samples is reached.  The result is the addition of five new
alternative training datasets for each fold of each language. The original test,
validation, and snippet test datasets are preserved, enabling direct comparison
between multi-language models trained on few-shot tasks and the \ac{sota}
single-language models.

\subsection{Models}
This study investigates the application of contemporary \ac{cnn}-based \ac{sota}
models for \emph{on-the-fly} \ac{sh} in both multi-language and few-shot learning
scenarios \cite{myPaper2}. The models used include three CNN-based variants with
increasing hidden unit sizes: \ac{cnn}32, \ac{cnn}64, and \ac{cnn}128. These
models represent the current benchmark for \ac{sh}, achieving the highest
accuracy on both valid and invalid language derivations while offering the
fastest inference times available \cite{myPaper2}. They set the standard for
single-language \ac{sh} tasks and were originally trained and validated on the
same dataset employed in this study, covering \java, \kotlin, \python, \cpp,
\csharp, and \javascript.

For this study, all models trained and validated on multi-language and few-shot
learning tasks are newly initialized instances of \ac{cnn}32, \ac{cnn}64, and
\ac{cnn}128. The architecture of these models follows the same design proposed by
Palma \etal, with only one modification: the size of the input layer.
In the work of Palma \etal~\cite{myPaper1, myPaper2}, the input size of each model was tailored to the
number of token types specific to the single language being evaluated. However,
the six languages considered in this study have varying token type counts. To
ensure feasibility, multi-language models require a fixed input dimension large
enough to accommodate the token types from all supported languages.
Additionally, the \ac{tn} strategy increases the input size slightly
to include a shared token region. Consequently, all multi-language models in this
study are configured with a uniform input size of 315.

To maintain consistency and eliminate any potential bias introduced by this
adjustment, single-language models have also been updated to the same input
dimension of 315. These models are retrained and validated on the exact
datasets, folds, and training configurations used in the original work by Palma
\etal. This adjustment ensures direct comparability of multi-language models with
their single-language \ac{sota} counterparts and validates that the evaluation
delays for multi-language models remain consistent with those of the
single-language models.
The resulting models are denoted as \emph{SL} models.

\subsection{Model Training}
All models evaluated in this study are trained using the same configuration
applied for \ac{sota} models \cite{myPaper1, myPaper2}. This configuration
specifies the optimizer, learning rate, batch size, and epoch count. Each model
is trained sequentially on the training samples using cross-entropy loss and the
Adam optimizer.

For all \ac{sh} tasks, including single-language, multi-language, and few-shot
scenarios, the training protocol consists of two epochs with an initial learning
rate of $10^{-3}$, followed by two additional epochs with the same learning rate
of $10^{-4}$~\cite{myPaper2}. This configuration is applied uniformly across all models,
including those employing \ac{tn}, ensuring consistency and comparability in the
training process.

\subsection{Scenarios}
\commentblock{
The experiments run in this work aim at the evaluation of the \ac{sh} accuracy
achievable though multi language models for each language considered and
compared these performances to the \ac{sota} single language models.
The training and validation processes of the \ac{sh} models considered in this
work are of two nature: Multi-Language Syntax Highlighting and Few-Shot
Multi-Language Syntax Highlighting. The rest of this section outlines how
validation tasks for models in these two categories are carried out in this work.
}
The experiments conducted in this study focus on evaluating the \ac{sh} accuracy
achievable by multi-language models for each language considered, comparing their
performance against \ac{sota} single-language models. The training and
validation processes for the \ac{sh} models evaluated in this work are divided
into two categories: \emph{Multi-Language Syntax Highlighting} and \emph{Few-Shot
Multi-Language Syntax Highlighting}. The remainder of this section details how
the validation tasks for models in these two categories are designed and
implemented.

\subsubsection{Multi-Language Syntax Highlighting}
Multi-Language \ac{sh} tasks involve training \ac{sh} models on datasets
containing multiple programming languages and evaluating the performance of the
resulting models on a per-language basis. These tasks leverage the
\emph{Multi-Language Dataset} and assess the performance of randomly initialized
\ac{cnn}32, \ac{cnn}64, and \ac{cnn}128 models.

For each of the three cross-validation folds of the \emph{Multi-Language
Dataset}, the models are trained on the training set of the respective fold
using the standard training procedure for \ac{sota} \ac{sh} models. This process
produces three multi-language models per fold, referred to as \emph{ML}32,
\emph{ML}64, and \emph{ML}128. For models incorporating \ac{tn}, the same
process is repeated using randomly initialized instances of \ac{cnn}32,
\ac{cnn}64, and \ac{cnn}128 with \ac{tn} enabled, resulting in models labeled as
\emph{ML32+TN}, \emph{ML64+TN}, and \emph{ML128+TN}.

All models trained in this task are evaluated on their \ac{sh} accuracy for each
of the six programming languages included in the dataset. Accuracy in this
context is defined as the percentage of non-whitespace tokens correctly
classified into their corresponding \ac{sh} class. Since the
\emph{Multi-Language Dataset} is constructed from per-language \ac{sh} oracles,
it represents the maximum achievable accuracy for \ac{sh} models in this task.
Additionally, \ac{sota} single-language models, which are \ac{cnn}-based
resolvers, achieve near-perfect accuracy and serve as the benchmark for
comparison \cite{myPaper1}.

Using the three-fold cross-validation split of the \emph{Multi-Language Dataset}
and evaluating models across all \emph{Coverage Tasks}, the resulting
\emph{ML\*} and \emph{ML+TN\*} models are assessed for their \ac{sh} accuracy on
both valid language derivation test sets and invalid language derivation test
sets (snippets). This evaluation ensures that per-language accuracy results can
be directly compared to the performance of single-language \ac{sota} resolvers,
providing a comprehensive analysis of the models' capabilities.

\subsubsection{Few-Shot Multi-Language Syntax Highlighting}
The \emph{Few-Shot Multi-Language} \ac{sh} tasks evaluate a model’s ability to
learn syntax highlighting patterns for previously unseen programming languages
when given only a limited number of training samples. Like the
\emph{Multi-Language Syntax Highlighting} tasks, these tasks involve training
models on datasets containing multiple programming languages and assessing their
performance on a per-language basis. However, unlike the full \emph{Multi-Language
Dataset}, the \emph{Few-Shot Multi-Language Dataset} is constructed by taking
subsets of varying sizes from the original training data, meaning models in
these tasks only have a limited number of samples—\emph{few-shots}—to learn syntax
highlighting patterns for each language.

The models evaluated in these tasks use the same architectures as those in the
\emph{Multi-Language} experiments, specifically the \emph{SL}32, \emph{SL}64, and
\emph{SL}128 variants. The \emph{Few-Shot} experiments follow a fine-tuning
approach: each model, originally trained on a single language, is fine-tuned on
a \emph{few-shot training subset} of the other five languages. For instance, a
\ac{cnn} model initially trained on \java is further fine-tuned using the
few-shot training data for \kotlin, \python, \cpp, \csharp, and \javascript.
This fine-tuned model is then evaluated on the validation datasets of each
language, allowing a direct comparison of its performance against both
single-language \ac{sota} models (\emph{SL\*}) and the multi-language models
(\emph{ML\*} and \emph{ML+TN\*}).

To maintain consistency and enable direct comparisons, the \emph{Few-Shot}
experiments utilize the same \emph{three-fold cross-validation splits} as those in
the \emph{Multi-Language} tasks. Additionally, the effect of increasing few-shot
sample sizes is analyzed by evaluating models trained on subsets of \emph{10, 30,
and 50} training samples per language. For each base language, a single
\emph{Few-Shot} experiment produces multiple models at different sample sizes. For
example, if Java is the base language of a \emph{SL}32, the fine-tuned models are denoted as
\emph{10-FS32-Java\*}, \emph{30-FS32-Java\*}, and \emph{FS-50-Java\*},
where the leading number indicates the few-shot sample size.

To evaluate the effectiveness of \ac{tn} in few-shot learning
scenarios, an additional set of experiments is conducted by replacing the base
model with a version that incorporates \ac{tn}. In these experiments, the
\ac{tn}-enabled model is fine-tuned on the same few-shot training subsets as the
standard models, ensuring a direct comparison between models with and without
token normalization. The \ac{tn} remains active throughout both the few-shot
training and validation phases.
Thus, following the previous example, the resulting models are denoted as \emph{10-FS32+TN-Java},
\emph{30-FS32+TN-Java*}, \emph{50-FS32+TN-Java}.
Evaluating the performance of these models enables an assessment of how \ac{tn}
enhances generalization in low-resource learning settings and whether it
improves accuracy when adapting to previously unseen programming languages with
limited training data.

This overall setup ensures a thorough evaluation of few-shot learning capabilities in
syntax highlighting and allows a direct comparison with both fully trained
\emph{single-language} models and \emph{multi-language} models trained on larger
datasets.

\section{Results}
\label{sec:results}

\subsection{RQ1 - Using Single-Language Models in Multi Language Tasks}

%
%
\begin{table}
\caption{
Average syntax highlighting accuracies on valid language derivations
for single-language SL models on their respective trained language (BASE) and
an unseen language (UNSEEN) across all coverage tasks.
}
\label{tab:rq1res}
\centering
\begin{tabular}{l l c c c c}
\toprule
~ & ~ & \multicolumn{1}{c}{T1} & \multicolumn{1}{c}{T2} & \multicolumn{1}{c}{T3} & \multicolumn{1}{c}{T4} \\
\midrule

\multirow{6}[2]{*}{BASE}~
    & SL32     & 99.65 & 99.65 & 99.37 & 99.37 \\
    & SL32+TN  & 99.66 & 99.64 & 99.37 & 99.38 \\
\cmidrule(lr){2-6}
    & SL64     & 99.67 & 99.66 & 99.41 & 99.41 \\
    & SL64+TN  & 99.67 & 99.67 & 99.41 & 99.42 \\
\cmidrule(lr){2-6}
    & SL128    & 99.67 & 99.67 & 99.42 & 99.42 \\
    & SL128+TN & 99.68 & 99.67 & 99.42 & 99.43 \\

\midrule

\multirow{6}[2]{*}{UNSEEN}~
    & SL32     & 40.60 & 40.28 & 38.04 & 36.80 \\
    & SL32+TN  & \tabhvalue 64.93 & \tabhvalue 63.56 & \tabhvalue 60.86 & \tabhvalue 59.69 \\
\cmidrule(lr){2-6}
    & SL64     & 44.70 & 41.76 & 39.53 & 36.95 \\
    & SL64+TN  & \tabhvalue 62.25 & \tabhvalue 60.75 & \tabhvalue 57.92 & \tabhvalue 57.43 \\
\cmidrule(lr){2-6}
    & SL128    & 43.55 & 41.64 & 38.60 & 37.18 \\
    & SL128+TN & \tabhvalue 61.10 & \tabhvalue 59.45 & \tabhvalue 57.71 & \tabhvalue 57.42 \\

\bottomrule
\end{tabular}
\end{table}

%
%
\begin{table}
\caption{
Average syntax highlighting accuracies on invalid language derivations
    for single-language SL models (\req{1}) on their respective trained language (BASE) and
an unseen language (UNSEEN).
}
\label{tab:rq1ressnip}
\centering
\begin{tabular}{l l c c c c}
\toprule
~ & ~ & \multicolumn{1}{c}{T1} & \multicolumn{1}{c}{T2} & \multicolumn{1}{c}{T3} & \multicolumn{1}{c}{T4} \\
\midrule
\multirow{6}[2]{*}{BASE}~
    & SL32     & 99.56 & 99.60 & 99.34 & 99.35 \\
    & SL32+TN  & 99.56 & 99.60 & 99.29 & 99.36 \\
\cmidrule(lr){2-6}
    & SL64     & 99.58 & 99.61 & 99.28 & 99.32  \\
    & SL64+TN  & 99.56 & 99.59 & 99.32 & 99.30 \\
\cmidrule(lr){2-6}
    & SL128    & 99.57 & 99.55 & 99.29 & 99.30 \\
    & SL128+TN & 99.59 & 99.61 & 99.31 & 99.38 \\

\midrule

\multirow{6}[2]{*}{UNSEEN}~
    & SL32     & 39.32 & 38.55 & 36.09 & 34.86 \\
    & SL32+TN  & \tabhvalue 62.83 & \tabhvalue 61.18 & \tabhvalue 58.44 & \tabhvalue 57.23 \\
\cmidrule(lr){2-6}
    & SL64     & 42.88 & 39.66 & 37.41 & 34.76 \\
    & SL64+TN  & \tabhvalue 60.19 & \tabhvalue 58.53 & \tabhvalue 55.49 & \tabhvalue 55.21 \\
\cmidrule(lr){2-6}
    & SL128    & 41.97 & 39.57 & 36.42 & 34.98\\
    & SL128+TN & \tabhvalue 58.67 & \tabhvalue 57.09 & \tabhvalue 55.32 & \tabhvalue 54.97 \\

\bottomrule
\end{tabular}
\end{table}

\req{1} investigates the performance of \ac{sota}
\ac{sh} models on unseen mainstream programming languages
and assesses the impact of \ac{tn} on their accuracy in
these scenarios. This evaluation is conducted by measuring the \ac{sh} accuracy
of \emph{SL} models trained following \ac{sota} standards
across various sizes, covering all six mainstream programming languages
considered in this study and previous research.

Following the experimental setup outlined in the experiments sections, \emph{Section} \ref{sec:experiments}, this study
evaluates the \ac{sh} accuracy of models \emph{SL32}, \emph{SL64}, and
\emph{SL128}, each trained on a specific programming language. Their accuracy is
assessed both on their trained language and on the remaining five languages, for
which they received no training. The evaluation employs three-fold
cross-validation, as organized in the \emph{Multi Language Dataset}, and
considers both valid language derivations and snippets. Additionally, the
performance of each model is analyzed for each \emph{CT}.
A similar procedure is conducted for models trained with the \ac{tn} feature
enabled, resulting in variants \emph{SL32+TN}, \emph{SL64+TN}, and
\emph{SL128+TN} for each of the six programming languages. These models undergo
the same three-fold cross-validation process to facilitate direct comparisons
with their \emph{SL} counterparts. This setup provides a comprehensive overview
of the \ac{sh} accuracy attainable by single-language models, with and without
\ac{tn}, across all \emph{CT}, both for their trained language and
for unseen languages.

Table \ref{tab:rq1res} presents the average \ac{sh} accuracy for \emph{SL*} and
\emph{SL*+TN} models when highlighting code in the language they were trained on
(denoted as \emph{BASE}) and when highlighting any of the five unseen languages
(denoted as \emph{UNSEEN}). The results confirm that \emph{SL} models achieve
near-perfect accuracy across all \emph{CT}s on their trained language, with
\ac{tn} having no significant effect in this case.
However, these models demonstrate poor generalization to unseen languages, with
average \ac{sh} accuracy reductions of: 57\% for \emph{T1}, 58\% for \emph{T2},
61\% for \emph{T3}, and 62\% for \emph{T4}.
The introduction of \ac{tn} improves accuracy in these cases, ensuring that
models can leverage their learned \ac{sh} logic from the trained language to
provide better performance on unseen languages. The observed improvements over
baseline \emph{SL*} models are consistent across tasks: 20\% improvement for
\emph{T1}, \emph{T2}, and \emph{T3}, and 21\% improvement for \emph{T4}.

Similarly, Table \ref{tab:rq1ressnip} reports the \ac{sh} accuracy for \emph{SL*} and
\emph{SL*+TN} models when highlighting invalid language derivations (i.e., code
snippets). These results mirror the trends observed for valid language
derivations. \emph{SL} models maintain near-perfect \ac{sh} accuracy for
snippets in their trained language, a consistency also observed in prior work~\cite{myPaper2}.
However, their accuracy drops even further when applied to unseen languages,
with an average accuracy loss between 58\% and 64\%, or more specifically:
58\% for \emph{T1}, 60\% for \emph{T2}, 63\% for \emph{T3}, and 64\% for \emph{T4}.
Once again, \ac{tn} mitigates these losses, improving accuracy for unseen
languages by: 19\% for \emph{T1}, 20\% for \emph{T2} and \emph{T3}, and 21\% for
\emph{T4}.

These results confirm that \emph{SL} models achieve near-perfect \ac{sh}
accuracy for the language they are trained on but are not generalizable to other
languages. This highlights the necessity for system integrators to train and
deploy separate \emph{SL} models for each programming language they wish to
support. However, enabling \ac{tn} significantly reduces accuracy losses on
unseen languages, with improvements of up to 21\%, suggesting that \ac{tn} is a
promising strategy for enhancing the generalization of syntax highlighting
models.

\subsection{RQ2 - Effectiveness of Multi-Language Models}
\begin{table*}
\caption{
Syntax highlighting accuracy results for valid language derivations across combinations of programming
language, coverage task, and model. The table compares multi-language
(\emph{ML*}) models (\req{2}) with \ac{sota} single-language (\emph{SL\*}) models (\req{1}).
Results include also variants for the \emph{SL\*} and \emph{ML\*} using token
normalization: \emph{SL\*+TN} and \emph{ML\*+TN}
Accuracy values are averaged across three-fold cross-validation and reported as
percentages.
}
\label{tab:req2res}
\tiny

\resizebox{\linewidth}{!}{
\begin{tabular}{
    l SSSS SSSS SSSS
}

\hiderowcolors
\toprule

\multirow{2}[2]{*}{\textbf{Model}} & \multicolumn{4}{c}{\textbf{\java}} & \multicolumn{4}{c}{\textbf{\kotlin}} & \multicolumn{4}{c}{\textbf{\python}} \\
\cmidrule(lr){2-5} \cmidrule(lr){6-9} \cmidrule(lr){10-13}
& {\textbf{\task{1}}} & {\textbf{\task{2}}} & {\textbf{\task{3}}} & {\textbf{\task{4}}} & {\textbf{\task{1}}} & {\textbf{\task{2}}} & {\textbf{\task{3}}} & {\textbf{\task{4}}} & {\textbf{\task{1}}} & {\textbf{\task{2}}} & {\textbf{\task{3}}} & {\textbf{\task{4}}} \\

\midrule
\showrowcolors
SL32
    & 99.95 & 99.93 & 99.89 & 99.88
    & 99.80 & 99.93 & 99.74 & 99.75
    & 100.00 & 99.89 & 99.90 & 99.89 \\

\rowcolor{gray!10}
SL32+TN
    & 99.95 & 99.92 & 99.88 & 99.89
    & 99.79 & 99.93 & 99.74 & 99.75
    & 100.00 & 99.89 & 99.89 & 99.90 \\

ML32
    & 99.91 & 99.88 & 99.82 & 99.83
    & 99.77 & 99.88 & 99.68 & 99.66
    & 100.00 & 99.88 & 99.87 & 99.86 \\

\rowcolor{gray!10}
ML32+TN
    & 99.93 & 99.91 & 99.87 & 99.86
    & 99.77 & 99.89 & 99.71 & 99.69
    & 99.99 & 99.89 & 99.89 & 99.89 \\

\midrule

SL64
    & 99.96 & 99.94 & 99.91 & 99.91
    & 99.79 & 99.94 & 99.76 & 99.76
    & 100.00 & 99.90 & 99.90 & 99.90 \\

\rowcolor{gray!10}
SL64+TN
    & 99.96 & 99.94 & 99.91 & 99.91
    & 99.81 & 99.94 & 99.75 & 99.76
    & 100.00 & 99.91 & 99.90 & 99.90 \\

ML64
    & 99.94 & 99.93 & 99.89 & 99.90
    & 99.80 & 99.92 & 99.73 & 99.73
    & 100.00 & 99.91 & 99.90 & 99.90 \\

\rowcolor{gray!10}
ML64+TN
    & 99.94 & 99.94 & 99.91 & 99.90
    & 99.79 & 99.92 & 99.74 & 99.74
    & 99.99 & 99.91 & 99.91 & 99.91 \\

\midrule

SL128
    & 99.97 & 99.94 & 99.92 & 99.91
    & 99.80 & 99.94 & 99.76 & 99.76
    & 100.00 & 99.91 & 99.91 & 99.91 \\

\rowcolor{gray!10}
SL128+TN
    & 99.96 & 99.95 & 99.91 & 99.91
    & 99.80 & 99.94 & 99.75 & 99.76
    & 100.00 & 99.91 & 99.91 & 99.91 \\

ML128
    & 99.95 & 99.94 & 99.91 & 99.91
    & 99.81 & 99.93 & 99.75 & 99.75
    & 100.00 & 99.92 & 99.91 & 99.92 \\

\rowcolor{gray!10}
ML128+TN
    & 99.95 & 99.94 & 99.92 & 99.91
    & 99.80 & 99.93 & 99.75 & 99.75
    & 99.99 & 99.92 & 99.92 & 99.92 \\

\hiderowcolors
\toprule

\multirow{2}[2]{*}{\textbf{Model}} & \multicolumn{4}{c}{\textbf{\cpp}} & \multicolumn{4}{c}{\textbf{\csharp}} & \multicolumn{4}{c}{\textbf{\javascript}} \\
\cmidrule(lr){2-5} \cmidrule(lr){6-9} \cmidrule(lr){10-13}
& {\textbf{\task{1}}} & {\textbf{\task{2}}} & {\textbf{\task{3}}} & {\textbf{\task{4}}} & {\textbf{\task{1}}} & {\textbf{\task{2}}} & {\textbf{\task{3}}} & {\textbf{\task{4}}} & {\textbf{\task{1}}} & {\textbf{\task{2}}} & {\textbf{\task{3}}} & {\textbf{\task{4}}} \\

\midrule
\showrowcolors

SL32
    & 98.68 & 99.33 & 98.25 & 98.26
    & 99.65 & 99.17 & 98.93 & 98.96
    & 99.83 & 99.64 & 99.49 & 99.50 \\

\rowcolor{gray!10}
SL32+TN
    & 98.68 & 99.33 & 98.24 & 98.25
    & 99.71 & 99.14 & 98.96 & 98.99
    & 99.82 & 99.64 & 99.50 & 99.52 \\

ML32
    & 98.44 & 99.12 & 97.90 & 97.93
    & 99.68 & 99.08 & 98.88 & 98.88
    & 99.82 & 99.63 & 99.47 & 99.45 \\

\rowcolor{gray!10}
ML32+TN
    & 98.56 & 99.21 & 98.06 & 98.05
    & 99.69 & 99.10 & 98.94 & 98.93
    & 99.80 & 99.64 & 99.45 & 99.43 \\

\midrule

SL64
    & 98.75 & 99.38 & 98.36 & 98.36
    & 99.69 & 99.16 & 98.99 & 99.00
    & 99.84 & 99.66 & 99.53 & 99.53 \\

\rowcolor{gray!10}
SL54+TN
    & 98.75 & 99.38 & 98.35 & 98.35
    & 99.67 & 99.20 & 99.02 & 99.05
    & 99.84 & 99.66 & 99.54 & 99.53 \\

ML64
    & 98.58 & 99.23 & 98.12 & 98.11
    & 99.73 & 99.16 & 98.98 & 98.97
    & 99.84 & 99.66 & 99.52 & 99.53 \\

\rowcolor{gray!10}
ML64+TN
    & 98.63 & 99.28 & 98.18 & 98.20
    & 99.72 & 99.17 & 98.99 & 99.00
    & 99.84 & 99.66 & 99.51 & 99.53 \\

\midrule

SL128
    & 98.77 & 99.40 & 98.40 & 98.40
    & 99.62 & 99.13 & 98.99 & 98.98
    & 99.84 & 99.67 & 99.54 & 99.54 \\

\rowcolor{gray!10}
SL128+TN
    & 98.77 & 99.40 & 98.40 & 98.40
    & 99.70 & 99.16 & 99.02 & 99.04
    & 99.85 & 99.67 & 99.54 & 99.54 \\

ML128
    & 98.64 & 99.27 & 98.20 & 98.20
    & 99.73 & 99.18 & 99.02 & 99.01
    & 99.85 & 99.67 & 99.55 & 99.55 \\

\rowcolor{gray!10}
ML128+TN
    & 98.67 & 99.31 & 98.24 & 98.24
    & 99.73 & 99.18 & 99.02 & 99.02
    & 99.84 & 99.67 & 99.55 & 99.54 \\

\bottomrule

\end{tabular}
}
\end{table*}

\begin{table*}
\caption{
Syntax highlighting accuracy results for invalid language derivations (code snippets)
across different programming languages, coverage tasks, and models. This table
compares multi-language (\emph{ML\*}) models (\req{2}) with \ac{sota} single-language
(\emph{SL\*}) models (\req{1}) and their variants using token normalization \emph{+TN}, reporting
accuracy values averaged over three-fold cross-validation.
}
\label{tab:req2ressnip}
\tiny

\resizebox{\linewidth}{!}{
\begin{tabular}{
    l SSSS SSSS SSSS
}

\hiderowcolors
\toprule

\multirow{2}[2]{*}{\textbf{Model}} & \multicolumn{4}{c}{\textbf{\java}} & \multicolumn{4}{c}{\textbf{\kotlin}} & \multicolumn{4}{c}{\textbf{\python}} \\
\cmidrule(lr){2-5} \cmidrule(lr){6-9} \cmidrule(lr){10-13}
& {\textbf{\task{1}}} & {\textbf{\task{2}}} & {\textbf{\task{3}}} & {\textbf{\task{4}}} & {\textbf{\task{1}}} & {\textbf{\task{2}}} & {\textbf{\task{3}}} & {\textbf{\task{4}}} & {\textbf{\task{1}}} & {\textbf{\task{2}}} & {\textbf{\task{3}}} & {\textbf{\task{4}}} \\

\midrule
\showrowcolors
SL32
    & 99.92 & 99.93 & 99.85 & 99.85
    & 99.74 & 99.93 & 99.69 & 99.69
    & 100.00 & 99.89 & 99.87 & 99.86 \\

\rowcolor{gray!10}
SL32+TN
    & 99.90 & 99.92 & 99.84 & 99.84
    & 99.74 & 99.93 & 99.69 & 99.69
    & 100.00 & 99.89 & 99.87 & 99.87 \\

ML32
    & 99.11 & 99.61 & 99.05 & 99.17
    & 99.50 & 99.64 & 99.37 & 99.17
    & 99.73 & 99.68 & 99.48 & 99.54 \\

\rowcolor{gray!10}
ML32+TN
    & 99.52 & 99.92 & 99.84 & 99.84
    & 99.05 & 99.93 & 99.69 & 98.97
    & 99.53 & 99.89 & 99.87 & 99.48 \\

\midrule

SL64
    & 99.91 & 99.95 & 99.89 & 99.88
    & 99.75 & 99.92 & 99.65 & 99.70
    & 100.00 & 99.88 & 99.88 & 99.89 \\

\rowcolor{gray!10}
SL64+TN
    & 99.91 & 99.95 & 99.89 & 99.90
    & 99.76 & 99.92 & 99.69 & 99.71
    & 100.00 & 99.89 & 99.88 & 99.89 \\

ML64
    & 99.79 & 99.78 & 99.84 & 99.73
    & 99.62 & 99.63 & 99.41 & 99.66
    & 99.85 & 99.76 & 99.76 & 99.78 \\

\rowcolor{gray!10}
ML64+TN
    & 99.89 & 99.94 & 99.88 & 99.85
    & 99.31 & 99.88 & 99.02 & 99.25
    & 99.72 & 99.83 & 99.56 & 99.63 \\

\midrule

SL128
    & 99.95 & 99.95 & 99.91 & 99.90
    & 99.76 & 99.88 & 99.68 & 99.71
    & 100.00 & 99.90 & 99.90 & 99.90 \\

\rowcolor{gray!10}
SL128+TN
    & 99.93 & 99.95 & 99.90 & 99.90
    & 99.76 & 99.93 & 99.70 & 99.64
    & 100.00 & 99.90 & 99.90 & 99.89 \\

ML128
    & 99.91 & 99.93 & 99.88 & 99.87
    & 99.10 & 99.60 & 99.16 & 99.04
    & 99.93 & 99.83 & 99.82 & 99.83 \\

\rowcolor{gray!10}
ML128+TN
    & 99.90 & 99.95 & 99.89 & 99.89
    & 99.75 & 99.91 & 99.70 & 99.70
    & 99.92 & 99.86 & 99.85 & 99.83 \\

\hiderowcolors
\toprule

\multirow{2}[2]{*}{\textbf{Model}} & \multicolumn{4}{c}{\textbf{\cpp}} & \multicolumn{4}{c}{\textbf{\csharp}} & \multicolumn{4}{c}{\textbf{\javascript}} \\
\cmidrule(lr){2-5} \cmidrule(lr){6-9} \cmidrule(lr){10-13}
& {\textbf{\task{1}}} & {\textbf{\task{2}}} & {\textbf{\task{3}}} & {\textbf{\task{4}}} & {\textbf{\task{1}}} & {\textbf{\task{2}}} & {\textbf{\task{3}}} & {\textbf{\task{4}}} & {\textbf{\task{1}}} & {\textbf{\task{2}}} & {\textbf{\task{3}}} & {\textbf{\task{4}}} \\

\midrule
\showrowcolors

SL32
    & 98.66 & 99.33 & 98.20 & 98.21
    & 99.24 & 99.17 & 99.02 & 99.05
    & 99.80 & 99.64 & 99.43 & 99.44 \\

\rowcolor{gray!10}
SL32+TN
    & 98.66 & 99.33 & 98.19 & 98.19
    & 99.29 & 99.14 & 98.70 & 98.70
    & 99.78 & 99.64 & 99.43 & 99.43 \\

ML32
    & 98.07 & 98.27 & 97.45 & 97.09
    & 99.15 & 99.18 & 98.57 & 98.74
    & 99.76 & 99.58 & 99.39 & 99.20 \\

\rowcolor{gray!10}
ML32+TN
    & 98.45 & 99.33 & 98.19 & 97.99
    & 98.79 & 99.14 & 98.70 & 98.44
    & 99.65 & 99.64 & 99.43 & 99.35 \\

\midrule

SL64
    & 98.75 & 99.36 & 98.29 & 98.30
    & 99.24 & 98.94 & 98.53 & 98.66
    & 99.80 & 99.62 & 99.45 & 99.46 \\

\rowcolor{gray!10}
SL64+TN
    & 98.74 & 99.35 & 98.28 & 98.28
    & 99.18 & 98.79 & 98.70 & 98.58
    & 99.79 & 99.62 & 99.47 & 99.46 \\

ML64
    & 97.78 & 98.06 & 96.93 & 97.26
    & 99.18 & 99.02 & 98.96 & 98.74
    & 99.79 & 99.61 & 99.45 & 99.46 \\

\rowcolor{gray!10}
ML64+TN
    & 98.57 & 99.22 & 98.11 & 98.09
    & 98.87 & 99.04 & 98.50 & 98.86
    & 99.66 & 99.63 & 99.38 & 99.48 \\

\midrule

SL128
    & 98.77 & 99.38 & 98.30 & 98.31
    & 99.17 & 98.55 & 98.51 & 98.49
    & 99.80 & 99.62 & 99.46 & 99.46 \\

\rowcolor{gray!10}
SL128+TN
    & 98.78 & 99.36 & 98.35 & 98.33
    & 99.24 & 98.91 & 98.53 & 99.06
    & 99.81 & 99.62 & 99.47 & 99.47 \\

ML128
    & 97.49 & 98.10 & 98.03 & 97.72
    & 99.33 & 98.61 & 98.94 & 98.74
    & 99.78 & 99.61 & 99.46 & 99.46 \\

\rowcolor{gray!10}
ML128+TN
    & 98.62 & 99.24 & 98.17 & 98.18
    & 98.99 & 98.83 & 98.60 & 98.73
    & 99.80 & 99.63 & 99.48 & 99.48 \\

\bottomrule

\end{tabular}
}
\end{table*}

\req{2} examines whether CNN-based architectures for \ac{sh} can be effectively
trained on multi-language datasets while maintaining comparable \ac{sh} accuracy
to \ac{sota} \emph{SL} models evaluated in \req{1}. The goal is to determine the
feasibility of multi-language training tasks without sacrificing accuracy, allowing
system integrators to minimize the number of \ac{sh} models deployed for fast
highlighting across multiple languages. Additionally, this evaluation
investigates the impact of multi-language training on \ac{sh} accuracy.

To assess this, the \ac{sh} accuracy of \ac{cnn}-based models is measured across
all six mainstream programming languages. The evaluation considers performance for
each \emph{CT} and compares these results with the \ac{sota} \emph{SL\*}
resolvers for each language and task. Accuracy values are averaged over a
three-fold cross-validation setup, following standard practices in the field.
This evaluation includes both valid and invalid language derivations or code
snippets. The multi-language models, denoted as \emph{ML32}, \emph{ML64}, and
\emph{ML128}, are trained according to the experimental setup detailed in
Section \ref{sec:experiments}, with separate models produced for each training
fold.

The \ac{sh} accuracy obtained for each combination of programming language,
\emph{CT}, and model is reported in Table \ref{tab:req2res}. The results
demonstrate that \emph{ML\*} models perform on par with their respective
single-language \emph{SL\*} counterparts, consistently achieving near-perfect
accuracy across all languages and tasks. For \emph{T1}, the average \emph{SL\*}
models achieve an accuracy of $99.67\% \pm 0.44$, while the \emph{ML\*} models
attain a nearly identical accuracy of $99.64\% \pm 0.49$. Similar results are
observed for \emph{T2} and \emph{T3}, with the most challenging task, \emph{T4},
yielding an average accuracy of $99.40\% \pm 0.58$ for \emph{SL\*} models and
$99.35\% \pm 0.64$ for \emph{ML\*} models.
Likewise, similar results are observed when highlighting invalid language
derivations, as presented in Table \ref{tab:req2ressnip}. This table, akin to
Table \ref{tab:req2res}, reports the average \ac{sh} accuracy for each combination of
model, \emph{CT}, and programming language.

Overall, these findings confirm that the high levels of accuracy achieved by
single-language models are also attainable with multi-language models. This
consistency across all languages and \emph{CT}s suggests that multi-language
training does not compromise \ac{sh} performance, making it a viable strategy
for real-world deployment.

\subsection{RQ3 - Effectiveness of Few-Shot Fine-Tuning for multi-language Syntax Highlighting}

%
%
\begin{table*}
\caption{
Synthesizing accuracy of all few-shot learning models per language and task. The
table includes models with and without the token normalizer (\ac{tn}), across
all model sizes (\emph{FS32}, \emph{FS64}, and \emph{FS128}) and few-shot
training sizes (10, 30, and 50 samples per language). Accuracy values are
reported for valid language derivations. The results reflect
the performance of models fine-tuned through few-shot learning on a previously
unseen target language while being trained on the other five languages. The
highest accuracy achieved for each combination of language, task, and few-shot
training size is highlighted.
}
\label{tab:rq34res}
\tiny

\resizebox{\linewidth}{!}{
\begin{tabular}{
    l cccc cccc cccc
}

\hiderowcolors
\toprule

\multirow{2}[2]{*}{\textbf{Model}} & \multicolumn{4}{c}{\textbf{\java}} & \multicolumn{4}{c}{\textbf{\kotlin}} & \multicolumn{4}{c}{\textbf{\python}} \\
\cmidrule(lr){2-5} \cmidrule(lr){6-9} \cmidrule(lr){10-13}
& {\textbf{\task{1}}} & {\textbf{\task{2}}} & {\textbf{\task{3}}} & {\textbf{\task{4}}} & {\textbf{\task{1}}} & {\textbf{\task{2}}} & {\textbf{\task{3}}} & {\textbf{\task{4}}} & {\textbf{\task{1}}} & {\textbf{\task{2}}} & {\textbf{\task{3}}} & {\textbf{\task{4}}} \\

\toprule

10-FS32     & 77.92 & 72.10 & 66.89 & 64.64 & 86.98 & 82.53 & 80.29 & 79.06 & 80.34 & 73.38 & 71.08 & 68.91 \\
\rowcolor{gray!10}
10-FS32+TN  & 83.80 & 81.16 & 76.54 & 75.18 & 90.22 & 87.99 & 86.02 & 84.93 & 83.61 & 82.59 & 80.41 & 79.48 \\
10-FS64     & 78.36 & 72.23 & 65.82 & 64.92 & 88.15 & 82.94 & 80.20 & 79.05 & 82.32 & 76.40 & 73.10 & 71.44 \\
\rowcolor{gray!10}
10-FS64+TN  & \tabhvalue 85.92 & 84.43 & \tabhvalue 80.80 & 78.80 & 91.45 & 89.62 & 87.36 & 85.93 & 86.30 & 85.15 & 83.71 & 82.42 \\
10-FS128    & 79.84 & 72.56 & 66.97 & 62.96 & 88.13 & 85.45 & 82.93 & 80.37 & 82.97 & 75.67 & 73.09 & 67.80 \\
\rowcolor{gray!10}
10-FS128+TN & 85.32 & \tabhvalue 84.85 & 79.88 & \tabhvalue 79.63 & \tabhvalue 91.97 & \tabhvalue 89.93 & \tabhvalue 87.66 & \tabhvalue 86.18 & \tabhvalue 89.02 & \tabhvalue 87.23 & \tabhvalue 85.07 & \tabhvalue 83.07 \\
\midrule
30-FS32     & 82.95 & 80.79 & 75.70 & 73.63 & 92.73 & 89.87 & 86.98 & 85.92 & 87.05 & 83.28 & 81.67 & 79.98 \\
\rowcolor{gray!10}
30-FS32+TN  & 89.92 & 89.25 & 85.90 & 84.46 & 94.43 & 92.61 & 90.89 & 89.91 & 92.44 & 91.69 & 90.07 & 89.79 \\
30-FS64     & 85.56 & 83.93 & 77.80 & 75.25 & 93.45 & 91.59 & 89.34 & 87.60 & 90.74 & 87.05 & 84.17 & 83.17 \\
\rowcolor{gray!10}
30-FS64+TN  & 92.33 & 92.26 & 90.30 & 89.18 & 95.91 & 94.43 & 92.94 & 91.96 & 95.28 & 93.43 & 93.05 & 92.33 \\
30-FS128    & 87.83 & 85.16 & 77.93 & 75.45 & 94.40 & 92.80 & 89.79 & 87.85 & 91.54 & 87.39 & 86.10 & 82.59 \\
\rowcolor{gray!10}
30-FS128+TN & \tabhvalue 93.28 & \tabhvalue 94.17 & \tabhvalue 91.07 & \tabhvalue 90.65 & \tabhvalue 96.14 & \tabhvalue 95.74 & \tabhvalue 93.23 & \tabhvalue 92.63 & \tabhvalue 96.39 & \tabhvalue 95.09 & \tabhvalue 94.55 & \tabhvalue 93.97 \\
\midrule
50-FS32     & 87.69 & 86.68 & 83.47 & 81.93 & 94.68 & 92.73 & 90.38 & 89.20 & 92.07 & 89.22 & 88.26 & 86.57 \\
\rowcolor{gray!10}
50-FS32+TN  & 94.16 & 93.48 & 91.56 & 90.72 & 96.20 & 94.79 & 93.22 & 92.63 & 95.94 & 94.85 & 93.75 & 93.59 \\
50-FS64     & 91.86 & 90.19 & 86.27 & 82.68 & 95.53 & 94.59 & 92.13 & 91.14 & 95.00 & 92.48 & 90.62 & 88.86 \\
\rowcolor{gray!10}
50-FS64+TN  & 95.91 & 95.14 & 94.07 & 93.56 & 97.11 & 96.24 & 94.73 & 94.20 & 97.34 & 96.14 & 95.81 & 95.68 \\
50-FS128    & 92.38 & 91.30 & 86.02 & 84.62 & 95.81 & 94.90 & 92.60 & 90.48 & 94.80 & 93.40 & 91.51 & 89.37 \\
\rowcolor{gray!10}
50-FS128+TN & \tabhvalue 96.44 & \tabhvalue 96.70 & \tabhvalue 95.37 & \tabhvalue 94.92 & \tabhvalue 97.25 & \tabhvalue 97.18 & \tabhvalue 95.42 & \tabhvalue 94.68 & \tabhvalue 97.98 & \tabhvalue 97.28 & \tabhvalue 96.65 & \tabhvalue 96.36 \\

\toprule

\multirow{2}[2]{*}{\textbf{Model}} & \multicolumn{4}{c}{\textbf{\cpp}} & \multicolumn{4}{c}{\textbf{\csharp}} & \multicolumn{4}{c}{\textbf{\javascript}} \\
\cmidrule(lr){2-5} \cmidrule(lr){6-9} \cmidrule(lr){10-13}
& {\textbf{\task{1}}} & {\textbf{\task{2}}} & {\textbf{\task{3}}} & {\textbf{\task{4}}} & {\textbf{\task{1}}} & {\textbf{\task{2}}} & {\textbf{\task{3}}} & {\textbf{\task{4}}} & {\textbf{\task{1}}} & {\textbf{\task{2}}} & {\textbf{\task{3}}} & {\textbf{\task{4}}} \\
\midrule

10-FS32     & 65.78 & 73.96 & 60.45 & 60.21 & 83.70 & 78.13 & 73.76 & 72.79 & 82.26 & 76.88 & 75.18 & 75.52 \\
\rowcolor{gray!10}
10-FS32+TN  & 71.19 & 77.07 & 64.49 & 65.95 & 87.13 & 83.83 & 80.27 & 80.18 & 86.26 & 83.80 & 81.69 & 81.74 \\
10-FS64     & 68.02 & 72.89 & 63.54 & 61.23 & 84.04 & 77.95 & 73.09 & 73.77 & 83.83 & 77.10 & 75.09 & 75.36 \\
\rowcolor{gray!10}
10-FS64+TN  & 78.93 & 82.46 & 73.78 & 72.77 & 87.74 & 83.78 & \tabhvalue 81.38 & 80.65 & 87.69 & 84.55 & 82.78 & 82.35 \\
10-FS128    & 72.02 & 78.78 & 66.25 & 64.04 & 84.28 & 79.23 & 72.52 & 72.23 & 84.05 & 77.15 & 75.42 & 75.01 \\
\rowcolor{gray!10}
10-FS128+TN & \tabhvalue 79.85 & \tabhvalue 85.98 & \tabhvalue 75.73 & \tabhvalue 76.55 & \tabhvalue 88.03 & \tabhvalue 85.08 & 81.00 & \tabhvalue 81.60 & \tabhvalue 89.22 & \tabhvalue 87.31 & \tabhvalue 84.64 & \tabhvalue 84.79 \\
\midrule
30-FS32     & 81.30 & 86.21 & 76.23 & 75.72 & 88.96 & 84.18 & 80.25 & 80.02 & 88.25 & 84.37 & 81.74 & 81.87 \\
\rowcolor{gray!10}
30-FS32+TN  & 85.35 & 90.48 & 80.90 & 81.29 & 92.53 & 89.71 & 86.63 & 86.88 & 91.51 & 90.19 & 88.39 & 88.73 \\
30-FS64     & 84.47 & 89.04 & 80.00 & 79.33 & 90.19 & 85.59 & 81.04 & 80.87 & 91.73 & 88.20 & 84.08 & 84.42 \\
\rowcolor{gray!10}
30-FS64+TN  & 88.88 & 92.78 & 85.58 & 85.44 & 93.69 & 90.99 & 89.04 & 88.66 & 94.74 & 93.18 & 91.45 & 91.02 \\
30-FS128    & 85.07 & 89.55 & 78.68 & 78.48 & 90.73 & 86.26 & 80.98 & 80.02 & 92.91 & 88.16 & 84.65 & 84.62 \\
\rowcolor{gray!10}
30-FS128+TN & \tabhvalue 89.41 & \tabhvalue 94.11 & \tabhvalue 86.26 & \tabhvalue 86.67 & \tabhvalue 94.33 & \tabhvalue 92.59 & \tabhvalue 89.41 & \tabhvalue 89.29 & \tabhvalue 95.10 & \tabhvalue 95.23 & \tabhvalue 92.81 & \tabhvalue 92.84 \\
\midrule
50-FS32     & 86.59 & 90.63 & 82.79 & 82.59 & 91.82 & 87.41 & 84.01 & 83.33 & 92.55 & 90.17 & 88.16 & 87.91 \\
\rowcolor{gray!10}
50-FS32+TN  & 89.90 & 94.31 & 87.04 & 87.39 & 94.36 & 92.67 & 91.01 & 91.18 & 94.16 & 92.70 & 91.48 & 91.85 \\
50-FS64     & 89.50 & 92.81 & 85.66 & 85.41 & 93.04 & 89.58 & 85.46 & 85.09 & 95.07 & 93.69 & 90.56 & 90.61 \\
\rowcolor{gray!10}
50-FS64+TN  & 92.69 & 95.66 & 90.47 & 90.17 & 95.70 & 94.51 & 92.82 & 92.40 & 96.61 & 95.60 & 94.37 & 94.38 \\
50-FS128    & 89.06 & 92.60 & 84.09 & 84.84 & 93.27 & 89.94 & 85.21 & 85.01 & 95.76 & 92.88 & 90.68 & 91.00 \\
\rowcolor{gray!10}
50-FS128+TN & \tabhvalue 93.10 & \tabhvalue 96.15 & \tabhvalue 90.62 & \tabhvalue 90.86 & \tabhvalue 96.14 & \tabhvalue 95.18 & \tabhvalue 93.39 & \tabhvalue 92.81 & \tabhvalue 96.95 & \tabhvalue 96.92 & \tabhvalue 95.17 & \tabhvalue 95.48 \\

\bottomrule
\end{tabular}
}
\end{table*}

\begin{table*}
\caption{
Synthesizing accuracy of all few-shot learning models per language and task. The
table includes models with and without the token normalizer (\ac{tn}), across
all model sizes (\emph{FS32}, \emph{FS64}, and \emph{FS128}) and few-shot
training sizes (10, 30, and 50 samples per language). Accuracy values are
reported for invalid language derivations. The results reflect
the performance of models fine-tuned through few-shot learning on a previously
unseen target language while being trained on the other five languages. The
highest accuracy achieved for each combination of language, task, and few-shot
training size is highlighted.
}
\label{tab:rq34ressnip}
\tiny

\resizebox{\linewidth}{!}{
\begin{tabular}{
    l cccc cccc cccc
}

\hiderowcolors
\toprule

\multirow{2}[2]{*}{\textbf{Model}} & \multicolumn{4}{c}{\textbf{\java}} & \multicolumn{4}{c}{\textbf{\kotlin}} & \multicolumn{4}{c}{\textbf{\python}} \\
\cmidrule(lr){2-5} \cmidrule(lr){6-9} \cmidrule(lr){10-13}
& {\textbf{\task{1}}} & {\textbf{\task{2}}} & {\textbf{\task{3}}} & {\textbf{\task{4}}} & {\textbf{\task{1}}} & {\textbf{\task{2}}} & {\textbf{\task{3}}} & {\textbf{\task{4}}} & {\textbf{\task{1}}} & {\textbf{\task{2}}} & {\textbf{\task{3}}} & {\textbf{\task{4}}} \\

\toprule
10-FS32     & 76.98 & 70.02 & 64.44 & 61.76 & 87.06 & 82.75 & 80.55 & 79.41 & 79.52 & 72.89 & 70.52 & 68.23 \\
\rowcolor{gray!10}
10-FS32+TN  & 83.37 & 80.38 & 76.39 & 74.64 & 90.58 & 88.66 & 86.26 & 85.40 & 83.10 & 82.10 & 79.96 & 79.11 \\
10-FS64     & 77.24 & 70.18 & 63.71 & 62.04 & 87.96 & 83.18 & 80.53 & 79.12 & 81.66 & 75.73 & 72.45 & 70.92 \\
\rowcolor{gray!10}
10-FS64+TN  & \tabhvalue 85.38 & 83.56 & \tabhvalue 79.82 & 77.88 & 91.76 & 90.25 & 87.95 & 86.40 & 85.73 & 84.72 & 83.25 & 81.93 \\
10-FS128    & 78.55 & 70.07 & 64.32 & 60.61 & 88.14 & 85.48 & 83.01 & 80.53 & 82.08 & 75.01 & 72.32 & 66.97 \\
\rowcolor{gray!10}
10-FS128+TN & 84.65 & \tabhvalue 83.93 & 79.17 & \tabhvalue 78.73 & \tabhvalue 92.12 & \tabhvalue 90.28 & \tabhvalue 88.01 & \tabhvalue 86.48 & \tabhvalue 88.64 & \tabhvalue 86.96 & \tabhvalue 84.69 & \tabhvalue 82.78 \\
\midrule
30-FS32     & 81.83 & 79.01 & 73.83 & 71.26 & 92.72 & 89.78 & 87.02 & 85.86 & 86.39 & 82.97 & 81.25 & 79.43 \\
\rowcolor{gray!10}
30-FS32+TN  & 89.34 & 88.35 & 84.93 & 83.10 & 94.63 & 92.92 & 91.04 & 90.07 & 91.97 & 91.35 & 89.76 & 89.44 \\
30-FS64     & 84.39 & 82.20 & 75.79 & 72.70 & 93.06 & 91.26 & 88.82 & 87.04 & 90.23 & 86.69 & 83.74 & 82.80 \\
\rowcolor{gray!10}
30-FS64+TN  & 91.50 & 91.24 & 89.16 & 87.83 & 95.99 & 94.78 & 93.08 & 92.11 & 94.85 & 93.12 & 92.75 & 91.92 \\
30-FS128    & 86.78 & 83.03 & 75.87 & 72.80 & 94.16 & 92.50 & 89.29 & 87.47 & 90.96 & 86.82 & 85.59 & 81.85 \\
\rowcolor{gray!10}
30-FS128+TN & \tabhvalue 92.70 & \tabhvalue 93.45 & \tabhvalue 90.21 & \tabhvalue 89.57 & \tabhvalue 96.14 & \tabhvalue 95.87 & \tabhvalue 93.24 & \tabhvalue 92.67 & \tabhvalue 96.06 & \tabhvalue 94.89 & \tabhvalue 94.34 & \tabhvalue 93.75 \\
\midrule
50-FS32     & 86.55 & 85.07 & 81.67 & 79.85 & 94.51 & 92.55 & 90.08 & 88.94 & 91.47 & 88.85 & 87.87 & 85.94 \\
\rowcolor{gray!10}
50-FS32+TN  & 93.49 & 92.65 & 90.52 & 89.47 & 96.27 & 94.99 & 93.28 & 92.65 & 95.56 & 94.59 & 93.49 & 93.33 \\
50-FS64     & 90.71 & 88.69 & 84.39 & 80.40 & 95.10 & 94.21 & 91.52 & 90.50 & 94.44 & 92.12 & 90.21 & 88.49 \\
\rowcolor{gray!10}
50-FS64+TN  & 95.17 & 94.29 & 92.99 & 92.34 & 97.15 & 96.40 & 94.77 & 94.24 & 96.97 & 95.90 & 95.55 & 95.36 \\
50-FS128    & 91.30 & 89.66 & 84.26 & 82.39 & 95.56 & 94.60 & 92.04 & 89.94 & 94.16 & 92.81 & 90.83 & 88.65 \\
\rowcolor{gray!10}
50-FS128+TN & \tabhvalue 95.96 & \tabhvalue 96.16 & \tabhvalue 94.71 & \tabhvalue 94.12 & \tabhvalue 97.21 & \tabhvalue 97.23 & \tabhvalue 95.37 & \tabhvalue 94.65 & \tabhvalue 97.62 & \tabhvalue 97.04 & \tabhvalue 96.38 & \tabhvalue 96.12 \\

\toprule
\multirow{2}[2]{*}{\textbf{Model}} & \multicolumn{4}{c}{\textbf{\cpp}} & \multicolumn{4}{c}{\textbf{\csharp}} & \multicolumn{4}{c}{\textbf{\javascript}} \\
\cmidrule(lr){2-5} \cmidrule(lr){6-9} \cmidrule(lr){10-13}
& {\textbf{\task{1}}} & {\textbf{\task{2}}} & {\textbf{\task{3}}} & {\textbf{\task{4}}} & {\textbf{\task{1}}} & {\textbf{\task{2}}} & {\textbf{\task{3}}} & {\textbf{\task{4}}} & {\textbf{\task{1}}} & {\textbf{\task{2}}} & {\textbf{\task{3}}} & {\textbf{\task{4}}} \\
\midrule

10-FS32     & 64.11 & 72.06 & 59.31 & 57.99 & 78.40 & 70.48 & 64.61 & 65.20 & 82.64 & 77.24 & 75.61 & 75.94 \\
\rowcolor{gray!10}
10-FS32+TN  & 70.32 & 75.32 & 62.77 & 64.81 & 82.28 & 76.14 & 71.07 & 70.86 & 86.90 & 84.04 & 82.04 & 82.07 \\
10-FS64     & 66.70 & 69.90 & 61.49 & 59.38 & 78.73 & 71.41 & 66.23 & 66.28 & 84.20 & 77.40 & 75.51 & 75.84 \\
\rowcolor{gray!10}
10-FS64+TN  & 77.88 & 81.14 & 72.59 & 71.00 & 81.87 & 75.67 & \tabhvalue 73.05 & 71.99 & 88.21 & 84.82 & 83.06 & 82.64 \\
10-FS128    & 71.05 & 77.35 & 65.53 & 63.70 & 79.96 & 73.42 & 67.13 & 67.10 & 84.19 & 77.25 & 75.62 & 75.11 \\
\rowcolor{gray!10}
10-FS128+TN & \tabhvalue 79.47 & \tabhvalue 84.84 & \tabhvalue 75.18 & \tabhvalue 75.79 & \tabhvalue 82.49 & \tabhvalue 77.60 & 72.69 & \tabhvalue 72.45 & \tabhvalue 89.67 & \tabhvalue 87.74 & \tabhvalue 85.06 & \tabhvalue 85.18 \\
\midrule
30-FS32     & 79.89 & 84.42 & 75.03 & 74.19 & 84.88 & 78.27 & 73.20 & 73.28 & 88.71 & 84.95 & 82.36 & 82.50 \\
\rowcolor{gray!10}
30-FS32+TN  & 84.21 & 89.22 & 79.74 & 80.14 & 90.84 & 85.24 & 80.48 & 82.43 & 92.09 & 90.71 & 88.92 & 89.28 \\
30-FS64     & 83.22 & 87.30 & 78.31 & 77.83 & 86.45 & 81.21 & 74.42 & 75.09 & 92.13 & 88.63 & 84.63 & 84.95 \\
\rowcolor{gray!10}
30-FS64+TN  & 88.33 & 91.49 & 84.66 & 84.51 & 90.15 & 85.77 & 84.92 & 83.07 & 94.94 & 93.53 & 91.80 & 91.40 \\
30-FS128    & 83.63 & 87.78 & 77.22 & 77.33 & 87.00 & 81.67 & 76.39 & 74.64 & 93.02 & 88.35 & 84.99 & 84.92 \\
\rowcolor{gray!10}
30-FS128+TN & \tabhvalue 88.61 & \tabhvalue 92.84 & \tabhvalue 85.30 & \tabhvalue 85.67 & \tabhvalue 91.72 & \tabhvalue 90.26 & \tabhvalue 85.08 & \tabhvalue 83.42 & \tabhvalue 95.24 & \tabhvalue 95.41 & \tabhvalue 93.22 & \tabhvalue 93.13 \\
\midrule
50-FS32     & 85.11 & 88.94 & 81.34 & 80.86 & 88.97 & 85.38 & 81.41 & 81.26 & 92.88 & 90.64 & 88.76 & 88.39 \\
\rowcolor{gray!10}
50-FS32+TN  & 89.11 & 93.14 & 86.37 & 86.53 & 93.69 & 91.62 & 89.65 & \tabhvalue 90.74 & 94.59 & 93.13 & 91.94 & 92.34 \\
50-FS64     & 88.52 & 91.05 & 84.15 & 83.91 & 91.73 & 86.90 & 83.95 & 83.06 & 95.23 & 93.96 & 90.86 & 90.88 \\
\rowcolor{gray!10}
50-FS64+TN  & 92.42 & 94.81 & \tabhvalue 89.92 & 89.58 & 94.57 & 93.47 & 91.34 & 90.38 & 96.72 & 95.86 & 94.55 & 94.64 \\
50-FS128    & 87.85 & 90.75 & 82.62 & 83.30 & 90.95 & 88.20 & 84.80 & 83.02 & 95.73 & 92.92 & 90.80 & 91.18 \\
\rowcolor{gray!10}
50-FS128+TN & \tabhvalue 92.50 & \tabhvalue 95.39 & 89.80 & \tabhvalue 89.83 & \tabhvalue 95.02 & \tabhvalue 94.43 & \tabhvalue 92.65 & 90.59 & \tabhvalue 97.04 & \tabhvalue 97.04 & \tabhvalue 95.41 & \tabhvalue 95.57 \\

\bottomrule
\end{tabular}
}
\end{table*}

\req{3} examines the accuracy of \ac{sh} achieved by
multi-language models trained with a limited number of examples per programming
language, a \emph{few-shot} approach. This differs from the \emph{ML\*} models,
which are trained on a comprehensive multi-language dataset comprising the union
of all single-language datasets. The goal of this investigation is to assess the
feasibility of training multi-language \ac{sh} models with reduced dataset
creation and training costs, particularly though few-show learning tasks.

The few-shot multi-language models (\emph{FS}) are trained following the
experimental setup outlined in Section \ref{sec:experiments}. Specifically,
\emph{FS32}, \emph{FS64}, and \emph{FS128} models are fine-tuned using
few-shot training sizes of 10, 30, and 50 samples per language fold.

The evaluation measures \ac{sh} accuracy of
\ac{cnn} models across six mainstream programming languages. The performance
is analyzed for each \emph{CT} and compared against both \ac{sota}
single-language (\emph{SL\*}) resolvers and the multi-language (\emph{ML\*})
models. Accuracy values are computed using a three-fold cross-validation setup,
as per standard practice in this domain, considering both valid and invalid
language derivations.

The results indicate that, regardless of the few-shot sample size or the base
language on which the model was pretrained, fine-tuning on a small sample set
produces multi-language models that outperform \emph{SL\*} models in multi-language
scenarios. As shown in Table \ref{tab:rq34res}, which reports the average
\ac{sh} accuracy of each \emph{FS} model per \emph{CT} and programming
language, fine-tuning with just 10 samples increases accuracy by 35\% compared to
the original non-finetuned model. Increasing the sample size to 30 and 50 leads
to further accuracy improvements of 45\% and 50\%, respectively. Similar
conclusions apply to \ac{sh} accuracy on invalid language derivations, as
detailed in Table \ref{tab:rq34ressnip}.

Additionally, the few-shot approach achieves superior \ac{sh} accuracy compared
to \emph{SL+TN} model variants, which leverage the \emph{TN} strategy for
enhanced language generalization. Few-shot models trained with 10, 30, and 50
samples yield accuracy improvements of 15\%, 24\%, and 29\% over \emph{SL+TN}
models. Likewise, for invalid language derivations, the same \emph{FS} models
achieve accuracy gains of 15\%, 25\%, and 31\%.

Despite reducing training samples and improving \ac{sh} performance in
multi-language scenarios over \emph{SL} and \emph{SL+TN} models, few-shot models
exhibit lower \ac{sh} accuracy than \emph{ML} models trained on 13,000
samples per language. The average accuracy gap between \emph{FS} and \emph{ML}
models is 24\%, 15\%, and 10\% for few-shot sizes of 10, 30, and 50, respectively.
Similar trends are observed for invalid language derivations, with accuracy
differences of 26\%, 16\%, and 11\%.
This means that although the few-shot learning approach is not capable
of replacing fully trained \emph{ML} models in outright \ac{sh} accuracy, it can boost the
accuracies achievable through \emph{SL} and \emph{SL+TN} models in multi-language
tasks, and it continues to outperform legacy \emph{state-of-practice}
resolvers such as Pygments and Tree-sitter~\cite{myPaper1,myPaper2}, which are widely used despite
substantially lower accuracy~\cite{myPaper1,myPaper2}. As such, few-shot models remain a viable and cost-effective upgrade
in scenarios where limited labeled data is available,
and where fully trained multi-language models are impractical.

\subsection{RQ4 - Impact of Token Normalization on Few-Shot multi-language Syntax Highlighting}

\req{4} investigates the effectiveness of \ac{tn} in enhancing
the \ac{sh} accuracy of multi-language models trained on a small number of
examples per programming language (\emph{few-shot}). The findings provide
insights into the feasibility of achieving multi-language \ac{sh} models at a
cost similar to \emph{FS} models through the application of \ac{tn}.
The few-shot models with \ac{tn} (\emph{FS+TN}) are trained following the
experimental setup outlined in Section \ref{sec:experiments}, resulting in
models \emph{FS32+TN}, \emph{FS64+TN}, and \emph{FS128+TN} for each training
fold and few-shot sample size (10, 30, and 50 examples per language). The
evaluation measures the \ac{sh} accuracy of \ac{cnn}-based models across six
mainstream programming languages, analyzing performance for each coverage task
and comparing results against baseline \emph{FS} models that do not incorporate
\ac{tn}. Accuracy values are averaged over a three-fold cross-validation setup,
consistent with standard practices for evaluating \ac{sh} per task and language.
Results account for both valid and invalid language derivations or snippets.

The results indicate that \ac{tn} consistently improves the accuracy of all
\emph{FS} models, irrespective of the few-shot training size. As detailed in
Table \ref{tab:rq34res} for valid language derivations and in Table
\ref{tab:rq34ressnip} for invalid derivations, \ac{tn} enhances the \ac{sh}
accuracy of every \emph{FS} model across all combinations of language, coverage
task, and training size.

For models trained on only 10 samples per language, \ac{tn} increases \ac{sh}
accuracy by an average of 8\% on both valid and invalid derivations. Models
trained on 30 samples per language experience a 6\% improvement in valid
derivations and a 7\% boost in invalid ones. Even at a training size of 50
samples per language, \ac{tn} maintains a positive impact, increasing accuracy
by an average of 5\% for both valid and invalid derivations.
Among all multi-language few-shot models, the use of the \ac{tn} yielded the
best performing model overall. This is the \emph{FS128+TN} model which
provides the highest \ac{sh} accuracy for any few-shot training size, and
the best overall model when operating on 50 few-shot training samples.
This configuration achieves the highest \ac{sh} accuracy among multi-language few-shot models.
This falls short of the near-perfect accuracy of single-language (\emph{SL}) models by an
average of $5\% \pm 1.20$ for valid language derivations and $6\% \pm 1.27$ for
invalid language derivations.

Another key observation is the increased consistency of \emph{FS+TN} models
across the four coverage tasks compared to their \emph{FS} counterparts. While
the \ac{sh} accuracy of baseline \emph{FS} models declines as the complexity of
the coverage task increases, this trend is not observed in the \emph{FS+TN}
variants. This suggests that \ac{tn} enables models to leverage similarities in
grammatical syntax across multiple languages, whereas baseline \emph{FS} models
must infer such patterns from limited training samples.

These results directly mitigate the accuracy tradeoff observed in few-shot multi-language models
without token normalization. While a performance gap relative to fully trained multi-language and
single-language models remains, the use of \emph{TN} substantially reduces this gap while preserving the
low training cost characteristic of few-shot learning. In practical terms, \emph{FS+TN} models approach the
accuracy of fully trained models far more closely than baseline FS models, while requiring only a
fraction of the training data. This makes FS+TN the preferred deployment strategy for practitioners
seeking a cost-effective solution, particularly in settings where training resources are constrained
or where rapid support for new or low-resource programming languages is required.
\emph{FS+TN} provides a favorable balance between accuracy and training cost, delivering
substantially higher precision than state-of-practice resolvers while avoiding the expense of fully
supervised multi-language training.

\subsection{Threats to Validity}
The state-of-practice resolvers for \ac{sh}, such as \emph{Pygments}~\cite{pygments} and
\emph{Tree-sitter}~\cite{treesitter}, which have been used as baselines in previous work,
support a significantly large number of programming languages. Notably,
\emph{Pygments} provides syntax highlighting for over 500 languages. A potential
limitation of this study is the evaluation of the proposed approach on a smaller
subset of languages: \emph{Java}, \emph{Kotlin}, \emph{Python}, \emph{C++},
\emph{C\#}, and \emph{JavaScript}.
While the selected languages exhibit substantial syntactic diversity and cover core constructs
shared by many mainstream programming languages, the generalization claims are inherently bounded by
the evaluated set. The proposed approach relies on what we term Deep Abstraction, where syntax
highlighting is learned by transpiling deterministic program behaviors into a statistical model,
rather than approximating unknown distributions. As such, the near-perfect token accuracy observed
suggests strong coverage of common syntactic patterns. However, languages with fundamentally
different grammatical structures, paradigms, or domain-specific syntax, such as systems programming
languages or highly specialized DSLs, may introduce patterns not represented in the current
evaluation. Assessing the behavior of the proposed models on such languages remains an open
direction and is left for future work to better characterize the limits of abstraction and
transferability.

Additionally, the experimental setup in this study focuses on multi-language and
few-shot models trained on all six languages included in the largest available
dataset for \emph{on-the-fly} \ac{sh}. However, the performance of these models
in scenarios involving more than six languages has not been investigated.
Expanding the evaluation to include a greater number of languages would provide
deeper insights into the scalability and potential limitations of the proposed
approach in handling diverse and larger multilingual datasets.

\section{Related Work}
\label{sec:related_work}

The primary motivation behind this work is to reduce the number of separately
deployed \ac{sh} models required by system integrators.
This challenge is addressed by the introducing \ac{multilang} models that replace existing
\ac{singlelang}, \ac{sota} solutions. In parallel, the training
overhead is also addressed by lowering the amount of data needed to produce accurate
\ac{sh} models—specifically, through a few‐shot learning
configuration and a token‐normalization strategy tailored to the highly
optimized, token‐based input these neural models expect.

These production and training overheads pose unique challenges not addressed
by prior \ac{multilang} model or
tokenization research. Existing approaches often assume large, flexible model
architectures and generalized token vocabularies~\cite{myPaper1, myPaper2}.
However, the specialized \ac{da} framework for \emph{on‐the‐fly} \ac{sh} relies on a tight
coupling to language‐specific integer tokens, allowing these models to run in
real time even under high request loads. The trade‐off is that typical
\ac{multilang} tokenization techniques do not directly apply, because they add
overhead and cannot leverage the strict, and minimally semantially valuable,
integer‐ID lexing that underpins fast inference.
With consideration of this field's specific requirements,
the next section reviews the most relevant methodologies in current literature,
highlighting how they compare to, and differ from, the approach proposed in this work.

\paragraph{Grammar‐Based and Rule‐Based Syntax Highlighters}
Early and widely adopted syntax highlighters, including \emph{Pygments} \cite{pygments}
and \emph{Tree-sitter}\cite{treesitter}, rely on extensive sets of regular expressions
or grammar rules that must be painstakingly maintained on a per‐language basis.
For instance, \emph{Pygments} already supports over 500 languages, however, developers
typically spend considerable time updating and revising these rules \cite{myPaper2}.
Similarly, \emph{Tree‐sitter} uses formal grammars for each language to produce accurate parse
trees. While these solutions are effective for many static use cases, they are
not suited to the \emph{on‐the‐fly} scenario because: they cannot gracefully handle
incomplete or invalid derivations, and they must either store vast libraries
of grammars or perform full parses under strict performance constraints.
Consequently, these approaches cannot easily address
the goal of one single model that automatically handles multiple languages and
partial code snippets in near real time.

\paragraph{Single‐Language Neural Syntax Highlighters}
Recent advances have replaced language‐specific highlighters with statistical or
neural models automatically compiled from brute‐force resolvers~\cite{myPaper1, myPaper2}.
In particular, \ac{cnn}-based methods achieve the highest inference performances
while retaining high coverage of each language’s syntax. This is accomplished
via \ac{da}, whereby a developer‐defined \ac{sh} oracle,
often expensive to build, is used to label large corpora; a \ac{cnn} then learns
the grammatical rules to replicate these labelling processes more efficiently then the
oracle, or \ac{bf} resolver. However, prior neural approaches remain
largely \ac{singlelang}: integrators must retrain new networks for each
language, thus facing substantial maintenance costs for \ac{multilang}
environments.

\paragraph{Large Multi‐Language Code Models}
Transformer‐based foundation models for
code, exemplified by
\emph{CodeBERT}~\cite{feng2020codebert},
\emph{CodeT5}~\cite{wang2021codet5},
\emph{PLBART}~\cite{ahmad2021unified},
\emph{UniXcoder}~\cite{guo2022unixcoder},
already
incorporate knowledge of multiple programming languages. They excel at tasks
such as code completion, search, summarization, and translation. Despite their
multi-language coverage, these models are typically large and computationally
expensive to train and run. In many cases, these modelas also rely on subword tokenization or different
embedding mechanisms that are not directly compatible with the carefully
minimized, integer-token input scheme required for \emph{on‐the‐fly} syntax
highlighting. Adapting these large models for real-time syntax
highlighting, especially when code may be partially invalid, poses a risk of lengthy
inference times and increasing system memory usage, making them less suitable
for the fast, token-ID centric pipelines that the problem statement in this work targets.

\paragraph{Multi‐Language Tokenization and Normalization}
Related studies on \ac{multilang} tokenization, such as unifying tokens across
languages for code transformation or \ac{multilang} code
search~\cite{feng2020codebert, guo2020graphcodebert, wang2021codet5, ahmad2021unified, salza2022effectiveness},
show that mapping common keywords or symbols onto shared embeddings can help a single
model generalize. However, most such approaches assume either open‐vocabulary
BPE/wordpiece methods~\cite{sennrich2015neural, kudo2018sentencepiece} or
uniform lexical boundaries for all languages—conditions
that do not hold in the specialized \ac{da} pipelines, which extract
only the integer token IDs from language‐specific lexers. Consequently, existing
multi‐language tokenizers cannot simply utilised without breaking the carefully
optimized input shape or the ability to handle invalid code fragments in a
robust, real‐time manner.

\paragraph{Few‐Shot Code Intelligence}
Recent work demonstrates that few‐shot or low‐resource code learning can be effective for tasks such
as code classification, summarization, or completion with minimal labeled data
~\cite{ahmed2022few, mann2020language, chen2021evaluating, lu2021codexglue}. These approaches typically
rely on large, pretrained transformer models, such as GPT‐3 or CodeT5, which can absorb cross‐language
syntax and vocabulary within multi‐billion parameter architectures and perform few‐shot inference via
prompting or brief fine‐tuning. However, while such methods achieve robust results with limited samples,
they are infeasible for \emph{on‐the‐fly} syntax highlighting scenarios: \textit{i}) the inference time
and memory requirements of large models can exceed practical limits for sub‐millisecond highlighting,
and \textit{ii}) subword tokenizers in these architectures conflict with the compact integer‐token
representations essential to \emph{deep‐abstracted}, \ac{cnn}‐based highlighters~\cite{myPaper2}.
Consequently, no existing few‐shot
techniques address a strict low‐latency context, where every millisecond matters and each token’s ID is
defined by a bespoke language‐specific lexer pipeline.

\section{Conclusions and Future Work}
\label{sec:conclusions}

On-the-fly \ac{sh} seeks to deliver fast, accurate highlighting of source code
in contexts where a traditional development environment is unavailable. Today’s
online software engineering tools frequently display or share code snippets and
full files in real time, underscoring the importance of highly efficient \ac{sh}
solutions. Achieving this goal relies on a \ac{da} approach that begins
with \ac{bf} syntax highlighters. Such \ac{bf} highlighters employ
a language’s lexer and parser to derive an \ac{ast}, from
which syntactic and grammatical tokens can be highlighted with maximum
precision. Although these \ac{bf} methods are computationally expensive, their
logic can be distilled and transferred into specialized neural models through a
carefully optimized input normalization process.

Historically, neural models derived from \ac{bf} highlighters have provided
near-perfect accuracy on both valid and invalid code derivations—a key strength
in online collaboration scenarios, where developers often display incomplete
code snippets or partially correct language constructs. However, two important
constraints limit the widespread adoption of this strategy: the substantial
effort required to collect large oracles of labeled data for every supported
programming language, and the need to deploy a specialized single-language
model for each language.
This work addresses both issues by introducing multi-language models for \ac{sh} that can cover
multiple languages, and by reducing training overhead through few-shot learning, which enables an
pre-trained single-language model to be incrementally adapted to new languages using only a small number of
additional oracle examples, rather than the approximately 13,000 samples previously required per
language.
The introduction of a
specialized token normalizer strategy further reduces the amount of training
data required, bolstering the viability of few-shot approaches.
These results demonstrate the viability of a single multi-language SH model that is fast through
consolidated deployment, adaptable through exposure to multiple languages, and both efficient and
scalable through the combined use of token normalization and few-shot learning.

While few-shot settings introduce a measurable accuracy gap relative to fully trained
models, the use of token normalization substantially mitigates this tradeoff by preserving the low
training cost characteristic of few-shot learning while bringing performance much closer to that of
fully trained \emph{ML} and \emph{SL} models. As a result, FS+TN configurations provide a
favorable balance between accuracy and cost, consolidating support for multiple languages into a
single deployed instance and offering a practical, cost-effective deployment strategy that
substantially exceeds state-of-practice syntax highlighting approaches, particularly in
resource-constrained and low-resource language scenarios.

At the same time, the token normalization routine proposed in this work exposes elements that
currently rely on systematic human evaluation. In particular, the construction of bindings for shared tokens
cannot be based solely on syntactic criteria such as rule definitions or regular expressions that
describe exact character-level matching. It also requires semantic judgment to determine whether
tokens that differ in their formal definitions nevertheless represent the same conceptual category
across languages. For instance, identifiers constitute a feature common to all programming
languages, yet the precise constraints on their form differ, such as whether identifiers may begin
with specific characters.  Future work should therefore explore the use of large language models to
support or automate these semantic decisions during the construction of token normalization
bindings. Unlike meta-learning approaches, this process does not require additional training data,
as the internal logic of token normalization is explicit and interpretable. Leveraging large
language models for semantic token alignment could eliminate the need for manual intervention,
enabling the normalization framework to scale more easily to additional languages while preserving
its efficiency and avoiding the training costs associated with meta-learning.

A promising direction for future work is the investigation of improved sampling strategies for
few-shot training in multi-language settings. In particular, ensuring that few-shot datasets include
representative examples of all major grammatical sub-productions of a target language could
substantially improve model generalization when only limited training data is available. This may be
achieved through informed sampling schemes or through automated sample generation based on the
formal grammatical definition of a language, enabling systematic coverage of syntactic constructs
with minimal data. Such approaches would reduce reliance on brute-force resolvers applied to large
corpora, lowering data generation costs while improving sample efficiency. This line of research is
especially relevant for cold-start scenarios and emerging programming languages, where sufficient
real-world samples may not yet exist, and could further narrow the remaining performance gap
observed in few-shot multi-language models.

Having resolved key challenges in training costs and deployment overhead, future
research should investigate the real-world impact of syntax-highlighting
accuracy delivered by these multi-language and few-shot models. While
near-perfect \ac{sh} is valuable in principle, its relative importance in
software engineering workflows, and the degree to which small accuracy trade-offs
are acceptable, garrants closer study. Such inquiries might look at whether minor
inaccuracies impact human performance in routine development tasks like code
reviews, code comprehension, or collaborative debugging. This, in turn, may
clarify the acceptable size of training sets and guide system integrators toward
informed trade-offs between model accuracy and resource expenditure.

Further investigation could also focus on how automated \ac{sh} has transformed
the tooling landscape. Shifting from extensive, developer-authored regular
expressions to specialized neural highlighters offloads complexity from human
experts onto a model that independently manages accuracy, coverage, and speed.
In addition to helping novice developers, this transition may increase the
proliferation of syntax highlighters across different programming languages or
domains. Future work should therefore examine how accessible this process is for
developers who have little experience with parser and lexer details. Moreover,
user studies can illuminate whether automated \ac{da} highlighters
reduce the need for deep language-grammar knowledge and whether they support
mainstream and new languages equally well. Finally, as these highlighters prove
increasingly robust in handling incorrect or partial language derivations,
evaluating their effectiveness in authentic online coding scenarios, ranging from
snippet-sharing platforms to real-time collaboration tools, will reveal the
degree to which improved accuracy influences overall software development
processes and collaboration practices.

\section*{Acknowledgements}

The research leading to these results has received funding from the Swiss National Science Foundation (SNSF) project \enquote{Melise - Machine Learning Assisted Software Development} (SNSF204632).

\balance
\bibliography{references, urls}

@inproceedings{myPaper1,
    author = {Palma, Marco Edoardo and Salza, Pasquale and Gall, Harald C.},
    title = {On-the-Fly Syntax Highlighting Using Neural Networks},
    year = {2022},
    isbn = {9781450394130},
    publisher = {Association for Computing Machinery},
    address = {New York, NY, USA},
    url = {https://doi.org/10.1145/3540250.3549109},
    doi = {10.1145/3540250.3549109},
    booktitle = {Proceedings of the 30th ACM Joint European Software Engineering Conference and Symposium on the Foundations of Software Engineering},
    pages = {269–280},
    numpages = {12},
    location = {<conf-loc>, <city>Singapore</city>, <country>Singapore</country>, </conf-loc>},
    series = {ESEC/FSE 2022}
}

@article{myPaper2,
  title={On-the-Fly Syntax Highlighting: Generalisation and Speed-ups},
  author={Palma, Marco Edoardo and Wolf, Alex and Salza, Pasquale and Gall, Harald C},
  journal={arXiv preprint arXiv:2402.08754},
  year={2024}
}

@inproceedings{10.1145/2858036.2858372,
    author = {Asenov, Dimitar and Hilliges, Otmar and M\"{u}ller, Peter},
    title = {The Effect of Richer Visualizations on Code Comprehension},
    year = {2016},
    isbn = {9781450333627},
    publisher = {Association for Computing Machinery},
    address = {New York, NY, USA},
    url = {https://doi.org/10.1145/2858036.2858372},
    doi = {10.1145/2858036.2858372},
    booktitle = {Proceedings of the 2016 CHI Conference on Human Factors in Computing Systems},
    pages = {5040–5045},
    numpages = {6},
    location = {San Jose, California, USA},
    series = {CHI '16}
}

@inproceedings{sarkar_impact_2015,
  title = {The {{Impact}} of {{Syntax Colouring}} on {{Program Comprehension}}},
  booktitle = {Annual {{Meeting}} of the {{Psychology}} of {{Programming Interest Group}} ({{PPIG}})},
  author = {Sarkar, Advait},
  year = {2015}
}

@article{feng2020codebert,
  title={Codebert: A pre-trained model for programming and natural languages},
  author={Feng, Zhangyin and Guo, Daya and Tang, Duyu and Duan, Nan and Feng, Xiaocheng and Gong, Ming and Shou, Linjun and Qin, Bing and Liu, Ting and Jiang, Daxin and others},
  journal={arXiv preprint arXiv:2002.08155},
  year={2020}
}

@article{guo2020graphcodebert,
  title={Graphcodebert: Pre-training code representations with data flow},
  author={Guo, Daya and Ren, Shuo and Lu, Shuai and Feng, Zhangyin and Tang, Duyu and Liu, Shujie and Zhou, Long and Duan, Nan and Svyatkovskiy, Alexey and Fu, Shengyu and others},
  journal={arXiv preprint arXiv:2009.08366},
  year={2020}
}

@article{wang2021codet5,
  title={Codet5: Identifier-aware unified pre-trained encoder-decoder models for code understanding and generation},
  author={Wang, Yue and Wang, Weishi and Joty, Shafiq and Hoi, Steven CH},
  journal={arXiv preprint arXiv:2109.00859},
  year={2021}
}

@article{ahmad2021unified,
  title={Unified pre-training for program understanding and generation},
  author={Ahmad, Wasi Uddin and Chakraborty, Saikat and Ray, Baishakhi and Chang, Kai-Wei},
  journal={arXiv preprint arXiv:2103.06333},
  year={2021}
}

@article{guo2022unixcoder,
  title={Unixcoder: Unified cross-modal pre-training for code representation},
  author={Guo, Daya and Lu, Shuai and Duan, Nan and Wang, Yanlin and Zhou, Ming and Yin, Jian},
  journal={arXiv preprint arXiv:2203.03850},
  year={2022}
}

@article{salza2022effectiveness,
  title={On the effectiveness of transfer learning for code search},
  author={Salza, Pasquale and Schwizer, Christoph and Gu, Jian and Gall, Harald C},
  journal={IEEE Transactions on Software Engineering},
  volume={49},
  number={4},
  pages={1804--1822},
  year={2022},
  publisher={IEEE}
}

@article{sennrich2015neural,
  title={Neural machine translation of rare words with subword units},
  author={Sennrich, Rico and Haddow, Barry and Birch, Alexandra},
  journal={arXiv preprint arXiv:1508.07909},
  year={2015}
}

@article{kudo2018sentencepiece,
  title={Sentencepiece: A simple and language independent subword tokenizer and detokenizer for neural text processing},
  author={Kudo, Taku and Richardson, John},
  journal={arXiv preprint arXiv:1808.06226},
  year={2018}
}

@article{mann2020language,
  title={Language models are few-shot learners},
  author={Mann, Ben and Ryder, N and Subbiah, M and Kaplan, J and Dhariwal, P and Neelakantan, A and Shyam, P and Sastry, G and Askell, A and Agarwal, S and others},
  journal={arXiv preprint arXiv:2005.14165},
  volume={1},
  pages={3},
  year={2020}
}

@article{chen2021evaluating,
  title={Evaluating large language models trained on code},
  author={Chen, Mark and Tworek, Jerry and Jun, Heewoo and Yuan, Qiming and Pinto, Henrique Ponde De Oliveira and Kaplan, Jared and Edwards, Harri and Burda, Yuri and Joseph, Nicholas and Brockman, Greg and others},
  journal={arXiv preprint arXiv:2107.03374},
  year={2021}
}

@inproceedings{ahmed2022few,
  title={Few-shot training llms for project-specific code-summarization},
  author={Ahmed, Toufique and Devanbu, Premkumar},
  booktitle={Proceedings of the 37th IEEE/ACM international conference on automated software engineering},
  pages={1--5},
  year={2022}
}

@article{lu2021codexglue,
  title={Codexglue: A machine learning benchmark dataset for code understanding and generation},
  author={Lu, Shuai and Guo, Daya and Ren, Shuo and Huang, Junjie and Svyatkovskiy, Alexey and Blanco, Ambrosio and Clement, Colin and Drain, Dawn and Jiang, Daxin and Tang, Duyu and others},
  journal={arXiv preprint arXiv:2102.04664},
  year={2021}
}

@online{pygments,
    title = {{Pygments}},
    author = {Brandl, Georg},
    url = {https://pygments.org},
    year = {2022}
}

@misc{treesitter,
  title = {Tree-sitter},
  author = {{Tree-sitter contributors}},
  year = {2024},
  note = {Version X.X.X},
  howpublished = {\url{https://tree-sitter.github.io/tree-sitter/}},
  url = {https://github.com/tree-sitter/tree-sitter},
}

@online{stackexchange,
    title = {{StackExchange Data Explorer}},
    author = {{Stack Exchange, Inc.}},
    url = {https://data.stackexchange.com},
    year = {2025}
}

@online{replicationpackage,
    title = {{Multi Language Models for On-the-Fly Syntax Highlighting}},
    author = {Palma, Marco Edoardo and Rani, Pooja and Gall, Harald C.},
    url = {https://doi.org/10.5281/zenodo.17266387},
    year = {2025}
}

\begin{complete-version}
\input{back/biography}
\end{complete-version}

\end{document}